# Nanoscale thermal imaging of dissipation in quantum systems

D. Halbertal[1], J. Cuppens[1,2], M. Ben Shalom[3], L. Embon[1,†], N. Shadmi[4], Y. Anahory[1], HR Naren[1], J. Sarkar[1], A. Uri[1], Y. Ronen[1], Y. Myasoedov[1], L. S. Levitov[5], E. Joselevich[4], A. K. Geim[3] and E. Zeldov[1]

**Energy dissipation is a fundamental process governing the dynamics of physical, chemical, and biological systems. It is also one of the main characteristics distinguishing quantum and classical phenomena. In condensed matter physics, in particular, scattering mechanisms, loss of quantum information, or breakdown of topological protection are deeply rooted in the intricate details of how and where the dissipation occurs. Despite its vital importance the microscopic behavior of a system is usually not formulated in terms of dissipation because the latter is not a readily measureable quantity on the microscale. Although nanoscale thermometry is gaining much recent interest[1-15], the existing thermal imaging methods lack the necessary sensitivity and are unsuitable for low temperature operation required for study of quantum systems. Here we report a superconducting quantum interference nano-thermometer device with sub 50 nm diameter that resides at the apex of a sharp pipette and provides scanning cryogenic thermal sensing with four orders of magnitude improved thermal sensitivity of below 1 μK/Hz$^{1/2}$. The non-contact non-invasive thermometry allows thermal imaging of very low nanoscale energy dissipation down to the fundamental Landauer limit[16-18] of 40 fW for continuous readout of a single qubit at 1 GHz at 4.2 K. These advances enable observation of dissipation due to single electron charging of individual quantum dots in carbon nanotubes and reveal a novel dissipation mechanism due to resonant localized states in hBN encapsulated graphene, opening the door to direct imaging of nanoscale dissipation processes in quantum matter.**

[1]Department of Condensed Matter Physics, Weizmann Institute of Science, Rehovot 7610001, Israel.
[2]Catalan Institute of Nanoscience and Nanotechnology (ICN2), CSIC and the Barcelona Institute of Science and Technology, Campus UAB, Bellaterra, 08193 Barcelona, Spain.
[3]National Graphene Institute, The University of Manchester, Booth St. E, Manchester M13 9PL, UK and the School of Physics and Astronomy, The University of Manchester, Manchester M13 9PL, UK.
[4]Department of Materials and Interfaces, Weizmann Institute of Science, Rehovot 7610001, Israel.
[5]Department of Physics, Massachusetts Institute of Technology, Cambridge, Massachusetts 02139, USA.
†Present address: Department of Physics, Columbia University, New York, New York 10027, USA.

Investigation of energy dissipation on the nanoscale is of major fundamental interest for a wide range of disciplines from biological processes, through chemical reactions, to energy-efficient computing[1-5]. Study of dissipation mechanisms in quantum systems is of particular importance because dissipation demolishes quantum information. In order to preserve a quantum state the dissipation has to be extremely weak and hence hard to measure. As a figure of merit for detection of low power dissipation in quantum systems[16] we consider an ideal qubit operating at a typical readout frequency of 1 GHz. Landauer's principle states the lowest bound on energy dissipation in an irreversible qubit operation to be $E_0 = k_B T \ln 2$, where $k_B$ is Boltzmann's constant and $T$ is the temperature[17,18]. At $T = 4.2$ K, $E_0 = 4 \times 10^{-23}$ J, several orders of magnitude below $10^{-19}$ J of dissipation per logical operation in present day superconducting electronics and $10^{-15}$ J in CMOS devices[19,20]. Hence the power dissipated by an ideal qubit operating at a readout rate of $f = 1$ GHz will be as low as $P = E_0 f = 40.2$ fW. The resulting temperature increase of the qubit will depend on its size and the thermal properties of the substrate. For example, a $120 \times 120$ nm$^2$ device on a 1 μm thick SiO$_2$/Si substrate dissipating 40 fW will heat up by about 3 μK (Fig. 1). Such signals are several orders of magnitude below the best sensitivity of several mK/Hz$^{1/2}$ of any of the existing imaging techniques[1-15] (Fig. 1a) including radiation-based thermometry utilizing infra-red[6], fluorescence in nanodiamonds[4,7,8], Raman[9], or transmission electron beam[5], and atomic force microscopy (AFM) outfitted with thermocouple or resistive thermometers[10-15]. Moreover, none of the existing imaging techniques was demonstrated to operate at sufficiently low temperatures that are essential for study of quantum systems.

Superconducting junctions are commonly used as highly sensitive thermometers[2,21], relying on the strong temperature dependence of their critical current $I_c(T)$, which in the vicinity of the critical temperature $T_c$ can be approximated by $I_c(T) \cong I_0(1 - T/T_c)$. Conventional junction configurations, however, are not suitable for scanning probe thermometry due to their planar geometry and strong thermal coupling to the substrate.
Here we introduce a non-contact cryogenic scanning probe microscopy technique based on the superconducting quantum interference device (SQUID) positioned on a tip[22,23]. We utilize a novel approach in which a single Pb ($T_c = 7.2$ K) superconducting junction or a SQUID is fabricated on the apex of a sharp quartz pipette (SQUID-on-tip (SOT)) which can be made as small as a few tens of nm in diameter (Supplementary Information S1). Figure 1b shows a scanning electron micrograph (SEM) of such a SOT thermometer (tSOT) with an effective diameter of 46 nm, as determined from its quantum interference pattern. Similar SOT devices have previously been reported[22,23] to be extremely sensitive magnetic sensors reaching magnetic spin sensitivity below 0.4 μ$_B$/Hz$^{1/2}$. The electrical characteristics of the tSOT (Fig. 1c) show the current through the sensor $I_{tSOT}$ vs.

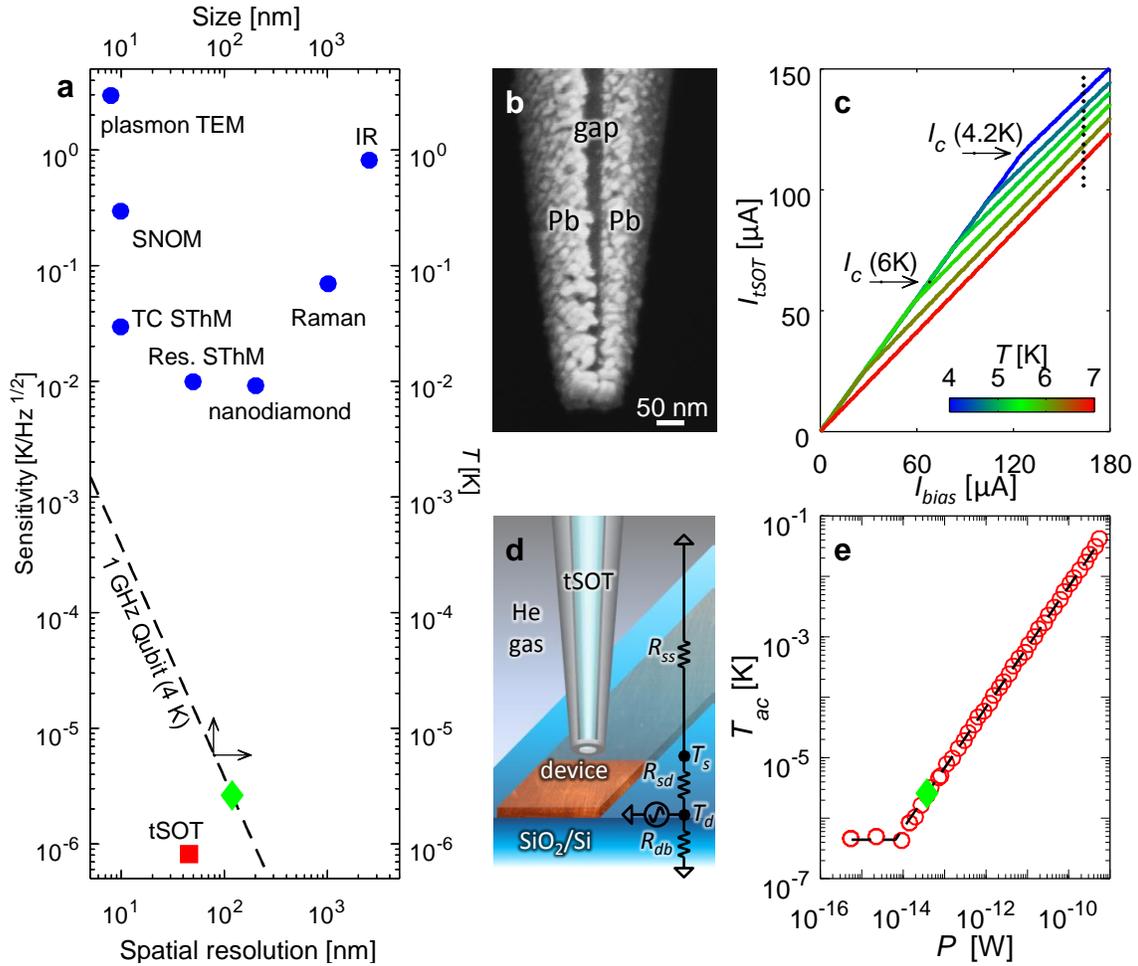

**Figure 1 | tSOT characteristics and performance. a**, Sensitivities of different thermal imaging techniques (blue) and of the tSOT (red) vs. their spatial resolution (bottom-left axes). Green diamond: the measured temperature increase due to 40 fW dissipation (taken from (e)) corresponding to Landauer's limit for qubit operation at 4.2 K at 1 GHz along with the theoretical scaling of the temperature with the qubit size (dashed line, top-right axes). **b**, SEM image of the 46 nm effective diameter Pb tSOT. **c,** Electrical characteristics of the tSOT for temperatures ranging from 4.2 K to 7.2 K with marked values of the critical current $I_c$ at representative temperatures. **d,** Schematic drawing of the measurement setup and a simplified effective thermal circuit. **e,** Measurement of the tSOT temperature $T_{ac}$ at 13.1 kHz above a 120 nm wide Cu nanowire with sheet resistance 0.46 Ω/□ vs. the *ac* power $P$ dissipated by a variable $I_{ac}$ in 120×120 nm² unit segment of the nanowire. The dashed line shows a linear fit (with slope $R_{db} = 6.8 \times 10^7$ K/W) and a noise floor of $T_{ac} \simeq 440$ nK below $P \simeq 6$ fW using lock-in amplifier time constant of 1 sec.

externally applied bias current $I_{bias}$ at various temperatures (see electrical diagram in Fig. S1c). For $I_{tSOT} < I_c$ essentially all the applied current flows through the tSOT while at higher bias a substantial part of the current diverts to a parallel shunt resistor. When biasing the tSOT at $I_{bias} > I_c$ (dashed line in Fig. 1c), the temperature dependence of $I_{tSOT}(T)$ gives rise to a thermal response $dI_{tSOT}/dT = -9.5$ µA/K (Fig. S1b), which combined with the very low white noise of $S_I^{1/2} = 8.3$ pA/Hz$^{1/2}$ of the tSOT (Fig. S1c) translates into a remarkably low thermal noise of $S_T^{1/2} = 870$ nK/Hz$^{1/2}$, a four orders of magnitude improvement over the existing thermal imaging methods (Fig. 1a).

To enable effective thermal imaging, the thermal properties of the sensor, including its coupling to the sample, are crucial. For non-invasive imaging, the thermal resistance between the sensor and the investigated device, $R_{sd}$, has to be significantly larger than the thermal resistance of the device to the bulk of its substrate, $R_{db}$ (Fig. 1d). On the other hand, in order for the temperature of the sensor $T_s$ to accurately describe the local temperature of the device $T_d$, a high thermal resistance $R_{ss}$ is required between the sensor and its support structure. The resulting overall requirement of $R_{ss} \gg R_{sd} \gg R_{db}$ is usually hard to achieve in AFM-type scanning thermal probes[12], leading to invasive in-contact imaging[13-14]. In tSOT, in contrast, $R_{ss}$ is extremely high, owing to its unique nanoscale cross-section geometry, giving rise to a quantum-limited phonon thermal conductivity[24], and to the absence of electronic heat conductivity along the superconducting leads. As a result, our smaller tSOT attains $R_{ss} \sim 10^{11}$ K/W (see S6) as compared to $R_{db} \sim 10^7$ K/W for a 120×120 nm$^2$ device on a SiO$_2$/Si substrate at 4.2 K (Fig. 1e). The corresponding optimal $R_{sd}$ of $10^8$ to $10^{10}$ K/W is readily achieved in our configuration using a few mbar of He exchange gas to tune $R_{sd}$ (Fig. S8). These features permit non-contact sensing of the local temperature $T_d$ of the sample with nanoscale resolution (Fig. S2 and sections S3-4).

To characterize the thermal sensitivity of the scanning tSOT we position the sensor above a 120 nm wide Cu nanowire on SiO$_2$/Si substrate (Figs. 1d and S7) carrying an *ac* current which results in an *ac* temperature modulation $T_{ac}$ of the nanowire at 13.1 kHz. By changing the current amplitude we measured tSOT $T_{ac}$ vs. the power $P$ dissipated per square 120×120 nm$^2$ segment of the wire (Fig. 1e). Since $T_{ac} \ll T_b$, where $T_b = 4.2$ K is the thermal bath temperature, the measurement is in the small signal limit and hence $T_{ac}$ is linear in $P$ as expected, reaching a noise level of $T_{ac} \simeq 440$ nK at $P \simeq 6$ fW. The green symbol in Fig. 1e points to the Landauer's dissipation limit of $P = 40.2$ fW of a qubit operating at 1 GHz at 4.2 K which gives rise to $T_{ac} = 2.6$ µK in our sample. This value is demarcated in Fig. 1a (green) along with the expected inverse scaling of the qubit temperature with its area (dashed line).

By applying a proper combination of magnetic field and bias, the tSOT can be tuned to have both magnetic field and thermal sensitivities. Since the Oersted field generated by the transport current in the sample is linear in current while the dissipation is quadratic, the resulting magnetic field signal $B_z^{ac}$ generated by the *ac* current $I_{ac}$ at frequency $f$ can be imaged simultaneously with the thermal signal $T_{ac}$ that will be present at frequency $2f$ as shown in Fig. S11.

We employed the tSOT technique for nanoscale thermal imaging of quantum matter. Figures 2a,b show $T_{ac}$ images of two single-wall carbon nanotubes (CNT) carrying an *ac* current of few nA. Each CNT is wound into a loop[25] as outlined by the dotted trajectories and shown by SEM images in Figs. 2d,e. The thermal signal in Fig. 2a tracks the CNT, revealing the current-driven dissipation along the entire CNT length. Surprisingly, and in contrast to the above, Fig. 2b shows an absence of heating in the circular part of the loop. The $T_{ac}$ image thus reveals that the applied current bypasses the loop exposing an electrically shorted junction between the two crossing sections of the CNT. This observation illustrates the capacity of tSOT for fault detection in operating nano-devices.

A striking feature evident in Figs. 2a,b is ring-like fine structures, with a zoomed-in $T_{ac}$ image of one of them shown in Fig. 2c. These features resemble the Coulomb blockade rings observed in scanning gate microscopy[26] in which the conductance through a quantum dot (QD) is measured as a function of the position of a conducting AFM tip. In this case, the characteristic equipotential rings corresponding to the periodic conditions of Coulomb blockade peaks governed by single electron charging of the QD[27] originate from the tip acting as a local gate. Since the tSOT is conducting, it can also serve as a nanoscale scanning gate. Our long CNTs on the $SiO_2$/Si substrate are, however, highly disordered, with numerous QDs in series resulting in CNT resistances in excess of 10 MΩ. As a result, no detectable change in the CNT conductance is observed while scanning the tSOT. The unique feature of the tSOT, however, is that in addition to functioning as a scanning gate it simultaneously operates as a nanoscale thermometer that detects minute changes in the local dissipation resulting from its own controllable local gating. The $T_{ac}$ ring in Fig. 2c with a cross sectional width of 20 nm (Fig. 2f) thus reveals μK-range changes of the temperature of a QD resulting from modulating its single electron Coulomb blockade conductance peaks by the scanning tSOT (see sections S13-14 of for additional details). This novel "scanning gate thermometry" thus adds another powerful functionality to the tSOT allowing nanoscale manipulation of the potential and study of the induced changes in the local scattering processes and dissipation that is inaccessible by other methods.

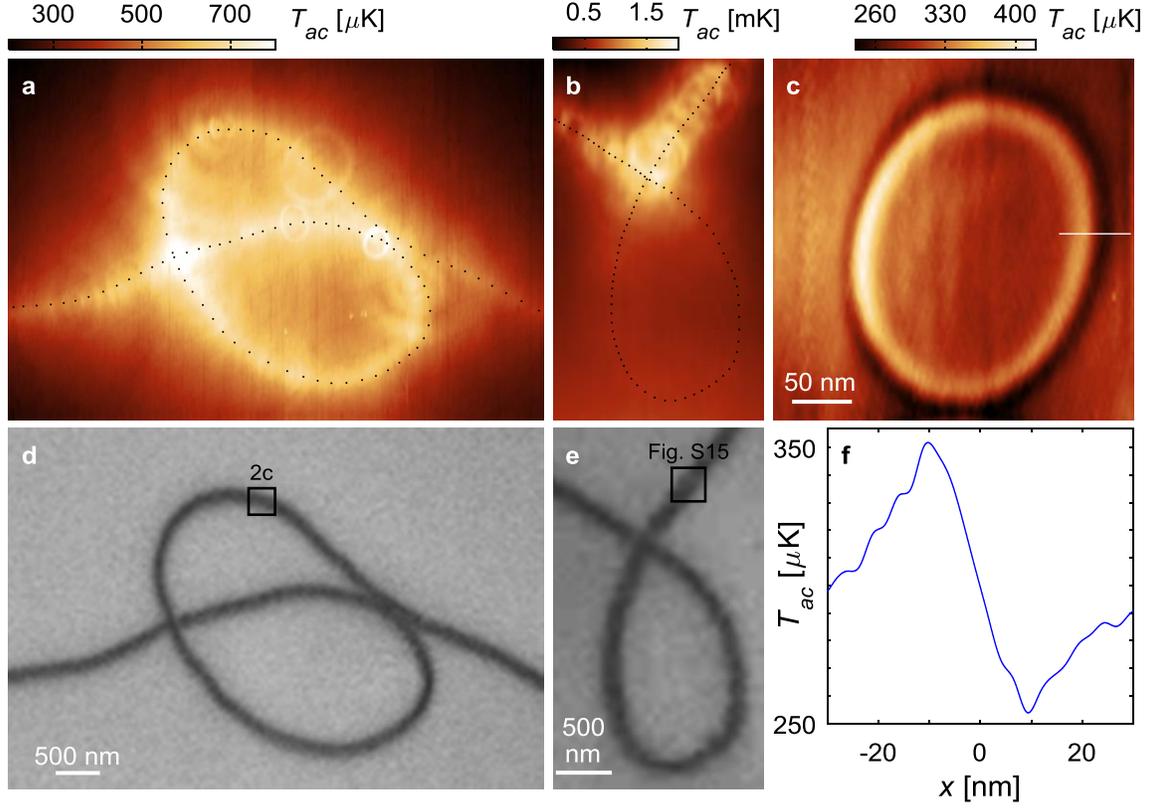

**Figure 2 | Thermal imaging of single-walled carbon nanotubes and scanning gate thermometry of quantum dots. a-b**, Thermal images of two CNT devices with loop geometry carrying $I_{ac}$ of 12 nA (a) and 3 nA (b) revealing an electrical short at the loop intersection point in (b). The $T_{ac}(x,y)$ was acquired by tSOT of 104 nm diameter at scanning heights of about 65 nm (a) and 150 nm (b). The ring-like structures in $T_{ac}$ result from variations in dissipation due to modulations in the single-electron charging of individual QDs as described in (c). **c**, Zoomed-in scanning gate thermometry image of a single QD in the area marked in (d) at a scan height of about 35 nm. Variations in the electrochemical potential of the QD induced by the scanning tSOT give rise to changes of the temperature of the QD at the Coulomb blockade peak conditions along the equipotential ring-like contour. **d-e**, SEM images of the devices in (a) and (b). **f**, Line-cut across the line marked in (c).

One of the topics of high interest in the studies of electronic transport at low temperatures is the non-equilibrium heating of the electron bath and proliferation of hot carriers due to poor electron-phonon coupling. Graphene offers a unique system where the typical distance for such carrier - lattice cooling can exceed the dimensions of the device[28] due to anomalously long lifetime of the hot carriers[29]. Electron cooling rates are believed to be sharply enhanced in the presence of disorder, however the precise mechanism of such enhancement is poorly understood[30].

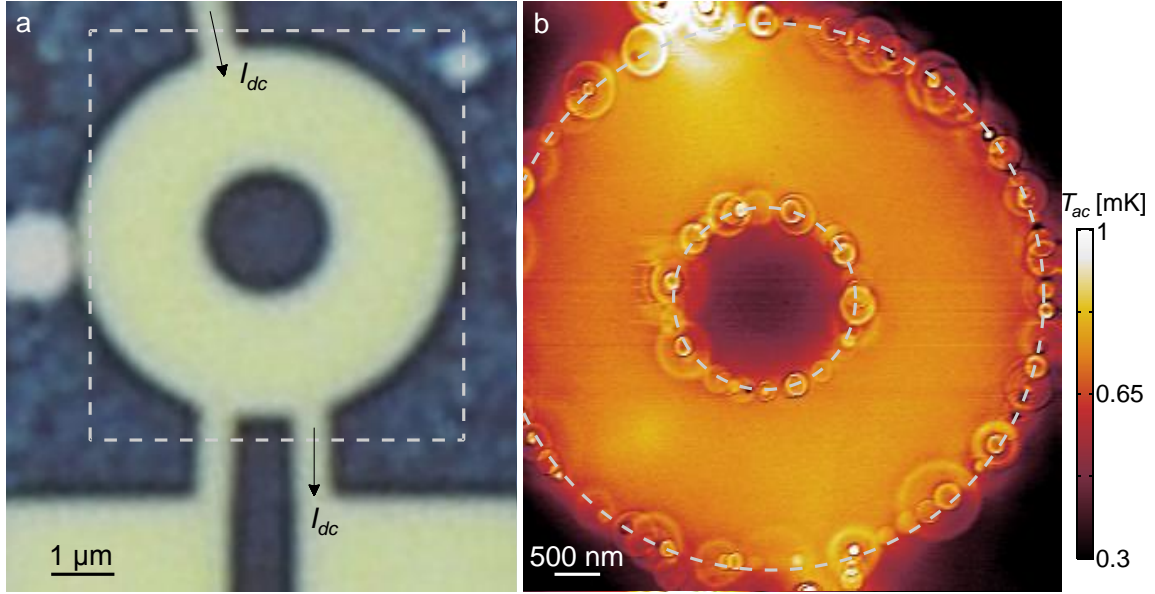

**Figure 3 | Scanning gate thermometry observation of dissipation at localized resonant states at graphene edges. a**, Optical image of hBN/graphene/hBN structure patterned into a washer shape (bright) with inner and outer diameters of 2 and 6 μm. **b**, Scanning *ac* gate thermometry $T_{ac}$ image of the outlined area in (a) in presence of a *dc* current of 6 μA applied as indicated by the arrows in (a) at carrier concentration of $10^{12}$ cm$^{-2}$. The $T_{ac}(x,y)$ was acquired by tSOT of 100 nm diameter at 4.2 K at a height of 40 nm in presence of an *ac* potential of 200 mV applied to the tip (see sections S13-S15). The dashed lines mark graphene edges. The necklace of rings reveals the presence of resonant states along the edges of graphene that serve as local centers of energy dissipation.

To probe microscopic dissipation mechanisms in graphene we performed scanning thermometry in high-mobility graphene encapsulated in hexagonal boron-nitride. We used a washer shaped device to which a *dc* current was applied between top and bottom constrictions as illustrated in Figure 3a (see S12 for details). Figure 3b shows the corresponding scanning gate thermal image revealing a startling visualization of dissipation processes, which are manifested in a complex structure of two necklaces of rings along the inner and outer edges of the device. The sharp ring-like structures have the same origin as the ring-like patterns in CNT in Figs. 2a-c, unveiling the presence of localized resonant states at the edges of graphene acting as atomic-scale heat beacons. The tSOT functions as a top gate which tunes the potential of the localized states into resonance when the combination of its distance and the voltage $V_{tSOT}$ applied between the tip and the sample match the resonance conditions. This gives rise to sharp rings of enhanced temperature as described in detail in sections S13 and S15. Consequently the ring radius $R$ changes upon varying $V_{tSOT}$ as demonstrated in Fig. S16 and Video S1 or upon changing the tip height as demonstrated in the case of CNT in Fig. S15.

Despite the similarity between the ring-like structures observed in CNT and graphene, their microscopic origin appears to be quite different. For CNT, the disorder potential creates QDs which confine discrete electronic states. Since the conductance of the QD is sharply enhanced under resonant conditions, the Coulomb blockade staircase gives rise to the sequence of concentric rings of enhanced local temperature in presence of current flow through the CNT (Fig. S15). In our hBN/graphene/hBN devices, in contrast, the hot electrons flow mainly in the bulk of the graphene, while the electron-lattice cooling occurs predominantly at defect states at graphene edges. Such defects can be formed by the vacancies and adatoms produced at the exposed graphene edge during etching process. Spatially localized electronic states originating from such defects, with energies pinned to the Dirac point[31] were recently revealed by scanning tunneling microscopy[32]. Localized states can significantly enhance electron-phonon coupling thus providing a local drain for the excess energy of the hot electrons[30]. Since each defect creates a single resonant electronic state[31] only one ring should be observed around each defect with the radius $R$ that depends on $V_{tSOT}$ as shown in Fig. S16 and Video S1. Acting as a gate-tunable spatially localized energy flow bottlenecks, these dissipation centers are likely to play a dominant role in the hot-carrier applications of graphene electronics. These observations shed new light on electron-lattice cooling mechanisms in graphene, a subject of high current interest.

The observation of striking spatially localized dissipation centers at graphene edge underscores the potential of the tSOT technique for uncovering the microscopic origins of dissipation in novel states of matter. Other systems of interest are topologically protected surface states, edge states in quantum spin and anomalous quantum Hall systems, and surface states in Weyl semimetals. By choosing appropriate superconducting materials[22,23,33] it should be possible to extend the operating temperature range of the tSOT from tens of mK to tens of K (Fig. S1d) thus allowing investigation of a wide range of quantum systems. In addition, the operation of the tSOT at elevated magnetic fields combined with the multifunctional abilities of magnetic sensing and scanning gate thermometry opens the way to nanoscale investigation and imaging of intricate thermoelectric and thermomagnetic phenomena including Nernst effect, thermal Hall effect, thermoelectric nanoscale cooling and quantum heat conductance.

**Acknowledgements**

We thank A. F. Young for discussions, M. V. Costache and S. O. Valenzuela for facilitation in Py/Cu sample fabrication, D. Shahar, I. Tamir, T. Levinson and S. Mitra for assistance in fabrication of a:$In_2O_3$ integrated devices, M. E. Huber for SOT readout setup, and M. L. Rappaport for technical assistance. This work was supported by the European Research Council (ERC) under the European Union's Horizon 2020 program (grant No 655416), by the Minerva Foundation with funding from the Federal German Ministry of Education and Research, and by Rosa and Emilio Segré Research Award. LSL and EZ acknowledge the support of MISTI MIT-Israel Seed Fund.

**Author contributions**

DH, JC and EZ conceived the technique and designed the experiments. DH and JC performed the measurements. DH performed the analysis and theoretical modelling. LE constructed the scanning SOT microscope. MBS and AKG designed and provided the graphene sample and contributed to the analyses of the results. NS and EJ fabricated the CNT samples. JC fabricated the Cu/Py sample. DH, HRN and JS fabricated the a:$In_2O_3$ sample. DH and YR designed and fabricated the spatial resolution demonstration sample. HRN, YA and AU fabricated the tSOT sensors. YA and YM developed the SOT fabrication technique. AU, YM and DH developed the tuning-fork based tSOT height control technique. LSL performed theoretical analysis. DH, JC and EZ wrote the manuscript. All authors participated in discussions and writing of the manuscript.

**Author information**

Correspondence and requests for materials should be addressed to DH (dorri.halbertal@weizmann.ac.il) or EZ (eli.zeldov@weizmann.ac.il).

# Supplementary Information:
# Nanoscale thermal imaging of dissipation in quantum systems

## S1. tSOT fabrication and thermal characterization

The tSOT fabrication methods, as well as the measurement circuitry and readout techniques, are based on the SOT technology[23,24,34] with several modifications. The tSOTs are fabricated by self-aligned three stage in-situ Pb deposition onto a pulled quartz pipette with apex diameter between 40 and 300 nm cooled to about 8 K. Pb has a very high surface diffusion and as a result grows in form of isolated islands if deposited at room temperature or even at 77 K. The low temperature deposition is therefore required in order to attain sufficiently uniform thin Pb films on the surface of the quartz pipettes as described in ref. 23. The two main modifications in tSOT fabrication are the use of grooved quartz tubes, which facilitate reliable fabrication of small tSOTs, and deposition of an integrated Au shunt resistor in the vicinity of the apex, which adds damping resulting in linear characteristics above $I_c$ (Fig. S1a) that are advantageous for thermal imaging. The current through the sensor $I_{tSOT}$ is measured using a SQUID series array amplifier (SSAA)[35] as shown in the inset of Fig. S1c. For $I_{tSOT} < I_c$ the device is in the superconducting state and essentially all the applied current $I_{bias}$ flows through the tSOT, while at higher currents the device is in a dissipative state in which $I_{bias}$ is distributed between the external shunt resistor $R_{shunt}$ and the tSOT, leading to a decrease in the slope of the characteristics.

Figure S1 shows the characteristics of the tSOT device used for the measurements in Figs. 1 of the main text. The tSOT effective diameter of 46 nm was extracted from the interference pattern as in Ref. 23. Figure S1a presents a set of 200 $I_{tSOT}$ vs. $I_{bias}$ curves at different temperatures $T$, color-coded between 4.2 K and $T_c = 7.2$ K, showing the monotonous decrease of $I_c$ with temperature. At a fixed $I_{bias} = 160$ µA (dotted line in Fig. S1a) this drop in $I_c$ results in a monotonous decrease in $I_{tSOT}(T)$, as shown in Fig. S1b, leading to a thermal response of $dI_{tSOT}/dT = $ -9.5 µA/K at the base temperature. Figure S1c presents the spectral noise density of $I_{tSOT}$, showing the typical $1/f$ noise followed by a white noise level of 8.3 pA/Hz$^{1/2}$ at frequencies above few hundred Hz. Combining the white noise level with the thermal response gives a thermal noise of 870 nK/Hz$^{1/2}$ at 4.2 K. Figure S1d shows the thermal noise vs. the temperature attained by measuring the noise spectra and the thermal response at various temperatures. This result demonstrates that the tSOT can operate over a wide temperature range, up to 7 K, with outstanding thermal sensitivity, down to 600 nK/Hz$^{1/2}$. Very close to $T_c$ the thermal noise diverges due to a decrease in the thermal response. Figure S1d shows that tSOT operates very well in the entire temperature range of 4.2 K to 7 K. In general, the most sensitive temperature range of the tSOT is $\sim T_c/2 < T < T_c$ where the critical current has an approximately linear $T$ dependence $I_c(T) \cong I_0(1 - T/T_c)$. At lower temperatures the $|dI_c(T)/dT|$ usually decreases and hence the sensitivity is expected to drop. This limitation can be readily addressed by using SOTs made of lower $T_c$ superconductors including Al [23] and In [33]. Another challenge for operation at very low temperatures is maintaining sufficiently low $R_{sd}$. Our preliminary studies show that In tSOT exhibits good thermal sensitivity in $^3$He exchange gas down to 700 mK. Reducing the working temperature further may be achieved by immersion in liquid $^3$He which should allow operation in $^3$He cryostat down to 250 mK and at even lower temperatures in a dilution fridge. For operation at higher temperatures NbN, MgB$_2$, and high-temperature superconductors could be potentially used, thus covering the full range of temperatures of interest for quantum systems.

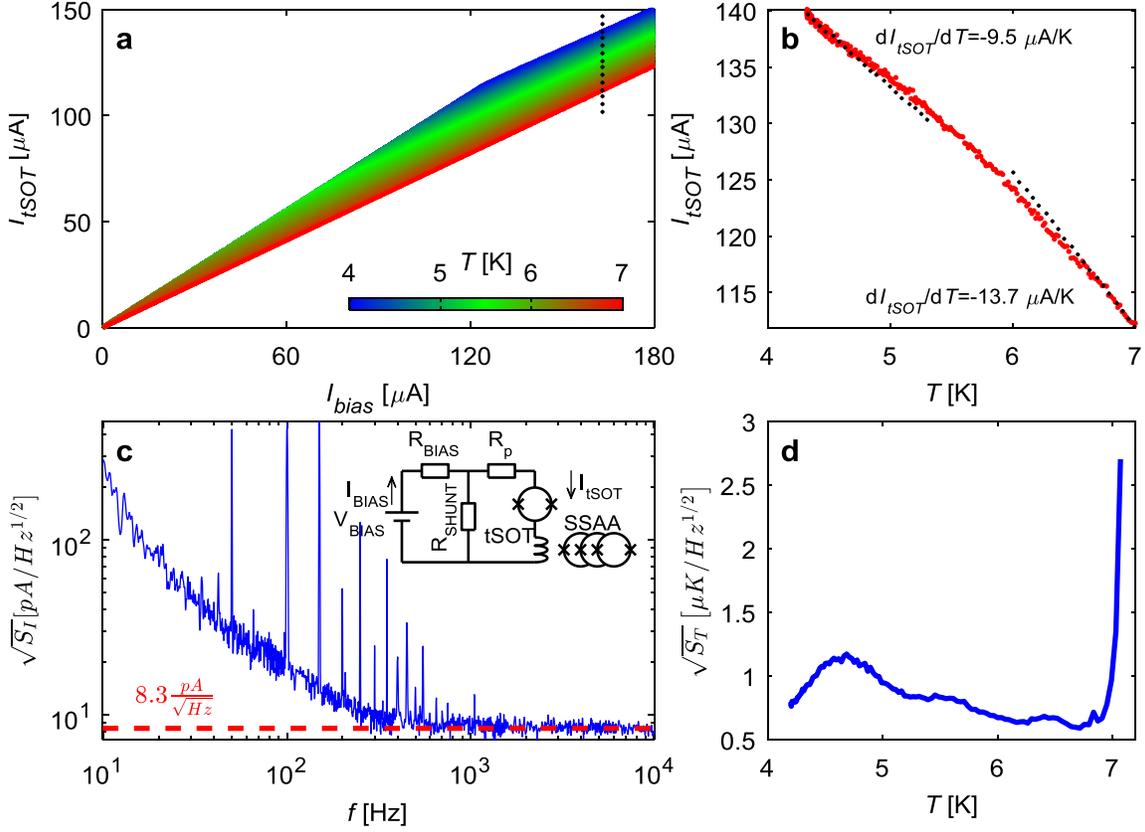

**Figure S1 | Characteristics of the 46 nm diameter tSOT. a**, Set of 200 $I_{tSOT}$ vs. $I_{bias}$ characteristics at various temperatures $T$ color coded from 4.2 to 7 K. **b**, Monotonically decreasing $I_{tSOT}$ vs. $T$ for $I_{bias} = 160$ µA giving rise to thermal response of −9.5 µA/K at the base temperature and −13.7 µA/K close to $T_c$. **c**, tSOT current noise spectral density at the working point at 4.2 K reaching a white noise level of 8.3 pA/Hz$^{1/2}$. Inset: schematics of the measurement circuit. **d**, Thermal noise as a function of temperature showing the very high thermal sensitivity of the tSOT in the entire range of temperatures up to 7 K.

## S2. Correlation between the sample and tSOT temperatures

Unlike most other scanning thermal microscopy (SThM) techniques, the tSOT is not in physical contact with the sample device under inspection. The heat mediation between the device and the sensor is carried by variable pressure He exchange gas. A dedicated multi-layer device (Fig. S2a) was fabricated in order to test the correlation between the temperature measured by the tSOT and the temperature of the device. The device comprised a SiO$_2$/Si substrate on which Au/Cr micro-heater (25 nm/5 nm thickness) was deposited and patterned into 12 µm wide strip (green in Fig. S2a), coated by 20 nm of Al$_2$O$_3$ insulating layer using atomic layer deposition (ALD). An amorphous In$_2$O$_3$ film[36] was deposited on top (33 nm thick, deposited in 1.5×10$^{-5}$ mtorr of O$_2$ and annealed in vacuum for 48 hours at 40° C) and patterned into an 8 µm wide strip with multiple contacts (red) connected to Au/Cr (25 nm/5 nm) metallic contact pads (blue).

The metal-insulator transition of the a:In$_2$O$_3$ gives rise to a strongly temperature dependent resistance $R(T)$ which we utilize for self-thermometry. By globally varying the system temperature $T$ we measure concurrently $R(T)$ of the device and $I_{tSOT}(T)$ of the tSOT under thermal equilibrium conditions between the device and the sensor as shown in Fig. S2b. From these data the corresponding thermal responses $dR/dT = -2.4$ kΩ/K and $dI_{tSOT}/dT = -10$ µA/K at 4.2 K are derived (Fig. S2b). Then, by driving a low frequency *ac* current in the Au micro-heater we induce local *ac* heating in the device and measure the resulting *ac* temperatures of the In$_2$O$_3$ film, $T_{ac}^d$, and of the tSOT sensor, $T_{ac}^s$, under steady state non-equilibrium thermal conditions.

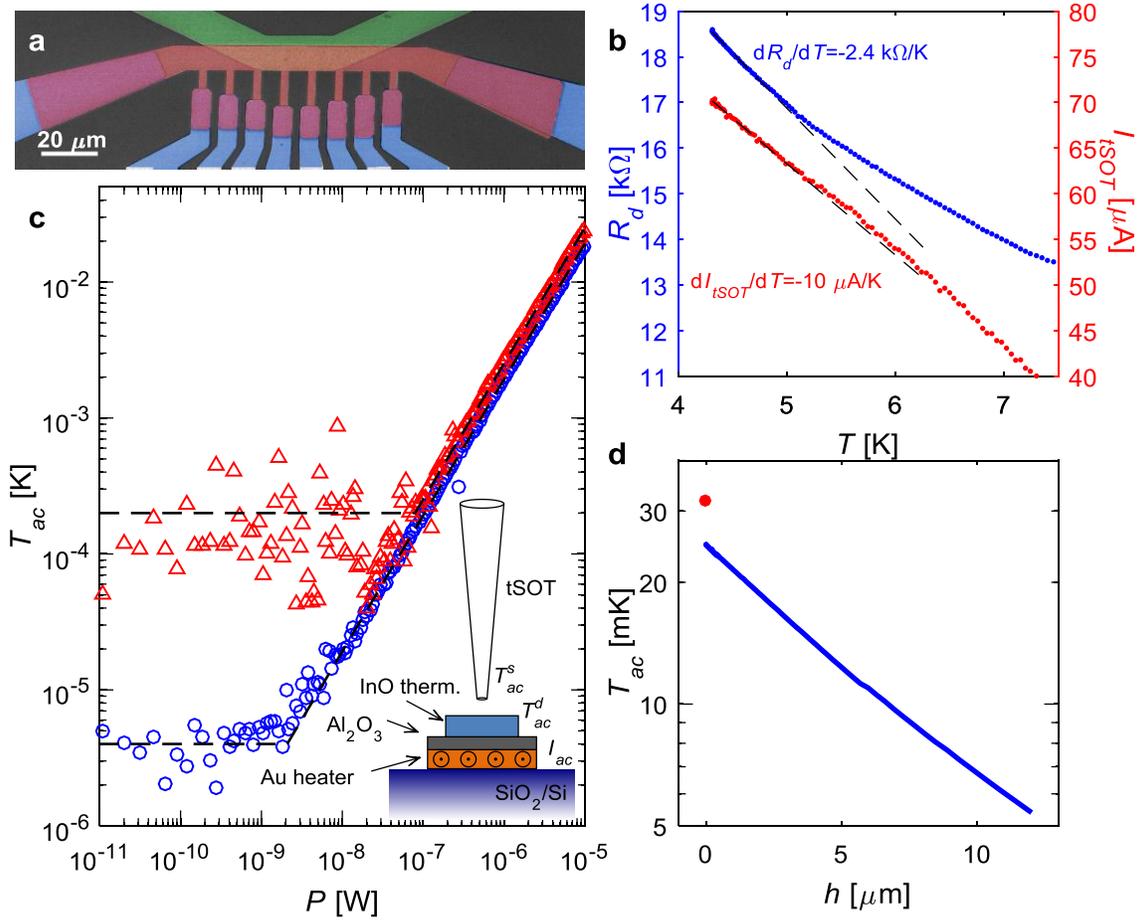

**Figure S2 | Comparison between the sample and tSOT temperatures. a,** False-color SEM image of the sample consisting of Cr/Au micro-heater (green), 33 nm thick a:In$_2$O$_3$ thin film thermometer with series of contacts (red), and Au/Cr leads (blue). The micro-heater is insulated from the a:In$_2$O$_3$ by 20 nm layer of Al$_2$O$_3$. **b,** Simultaneous four probe measurement of the resistance $R(T)$ of the central In$_2$O$_3$ segment of the device (blue) and of the current $I_{tSOT}(T)$ through the tSOT (red) vs. temperature $T$ under thermal equilibrium conditions. The dashed lines show the corresponding slopes and the thermal responses at 4.2 K. **c,** Simultaneously measured second-harmonic $T_{ac}^d$ of the In$_2$O$_3$ thermometer and $T_{ac}^s$ of the tSOT vs. the total power $P$ dissipated in the Au heater by $ac$ current at 10 Hz demonstrating the close conformity of the local temperature measurement of the tSOT. Dashed lines are linear fits to the data and demarcate the thermal noise levels of the two measurements. Inset: Schematic cross section of the sample and of the tSOT positioned at a relatively large height of 500 nm above the central segment of the In$_2$O$_3$ film. **d,** $T_{ac}^s$ as a function of tSOT height $h$ above the device (blue) and $T_{ac}^d$ of the central segment of the device (red).

Imaging was done, here and elsewhere, using an in-house-built scanning microscope operating at 4.2 K as described in ref. 23. The measurement was performed by driving an *ac* current of variable amplitude up to 0.35 mA at $f = 10$ Hz to the underlying Au micro-heater of 142 Ω resistance. The use of a low frequency is a result of the high RC time constant of the a:In$_2$O$_3$ device due to its high resistance (2-probe resistance of 200 kΩ). The resulting *ac* temperature variation of the device $T_{ac}^d$ at $2f$ was measured by applying a probing *dc* current of 1 μA to the current leads of the In$_2$O$_3$ device and measuring the second-harmonic voltage signal on the central segment of the device using a lock-in amplifier. A tSOT of 110 nm diameter and 2.1 μK/Hz$^{1/2}$ temperature sensitivity (in the white noise region above few hundred Hz) was positioned 500 nm above the center of the central segment of the device (Fig. S2c inset) in 16 mbar of He exchange gas. The relatively large height of the tSOT was used for convenience since only the average local temperature measured by the tSOT was of importance for direct comparison with the temperature reading of the device. Figure S2c shows the simultaneously measured $T_{ac}^d$ and $T_{ac}^s$ vs. the total power $P$ dissipated in the micro-heater. The dashed lines show a linear fit to the data and also mark the noise levels of 200 μK and 4 μK of the two measurements using 1 sec time constant of the lock-in amplifiers. The two signals scale linearly with $P$ as expected with $T_{ac}^s$ closely following $T_{ac}^d$ thus establishing a measurement of the actual device temperature by the tSOT.

Next we applied 0.35 mA *ac* current in the heater and measured the $T_{ac}^s$ of the tSOT vs. the height $h$ above the device surface up to $h = 12$ μm. Figure S2d shows an exponential decay of $T_{ac}^s$ with the height with a characteristic length scale of 7 μm. Similar results were attained for He gas pressures in the range of 2 to 26 mbar. Interestingly, on approaching $h = 0$ the data show a higher $T_{ac}^d$ by about 22% than $T_{ac}^s$. A similar 24% systematic difference is also observed in the data of Fig. S2c (that was taken 500 nm above the device). Such a difference can be readily accounted for by taking into account the thermal Kapitza resistance $R_K$ at the In$_2$O$_3$/He gas interface which leads to temperature discontinuity at the solid/gas interface due to phonon mismatch[37]. Considering the measured temperature discontinuity and an estimate for the heat flow to the gas in our geometry we attain $R_K \cong 6$ K·cm$^2$/W which is consistent with the values of the Kapitza resistance between solids and helium gas at 4.2 K available in the literature[38].

## S3. Theoretical analysis of the tSOT spatial resolution

In the presence of local dissipation we define $\delta T(x,y)$ to describe the spatially varying surface temperature of the sample taken relative to the bath temperature $T_0$. Without loss of generality we will work in the limit $\delta T(x,y) \ll T_0$. The corresponding temperature $\delta T_{tSOT}(x,y)$ measured by the tSOT (relative to $T_0$) will be determined by a convolution of $\delta T(x,y)$ with a heat transfer kernel $W(r)$, where $r$ is the radial distance from the tip center. The effective spatial resolution is then determined by the kernel $W(r)$ which depends on the diameter $D$ of the tip and the scanning height $h$ above the surface.

Our analysis is based on the assumption of ballistic flow of He atoms between the sample surface and the tSOT which for our geometry holds for He gas pressures of up to about 50 mbar [39]. We assume for simplicity that each atom carries an excess energy proportional to the surface temperature $\delta T(x,y)$ at the point of its origin and deposits this entire excess energy to the tSOT upon first encounter with its surface. We also assume that He atoms are scattered from the sample surface with a uniform angular distribution into the half space above the surface. These assumptions make our estimate a worst case scenario, while a more complete analysis should predict a better resolution. This is because the scattering probability is expected to be higher for smaller azimuthal angles and also because we ignore the contributions of multiple atom collisions between the sample and the tSOT which are more likely to happen for atoms trapped in the narrow gap between the tip apex and the sample.

We model the tSOT as a disc of diameter $D$ that resides in the $z = h$ plane, centered at $x = y = 0$. A planar sample is positioned at the $z = 0$ plane. Following the above assumptions the probability density distribution of atoms departing from the origin within a small solid angle $\Delta \Omega$ that covers an area $\Delta S$ on the $z = h$ plane positioned at a radial distance $r$ is $\Delta \Omega / \Delta S = h(h^2+r^2)^{-3/2}$. Using this expression we estimate the probability of an atom originating at $\boldsymbol{r} = (x, y)$ on the sample surface to collide with the tSOT. The resulting 2D heat transfer kernel $W(\boldsymbol{r})$ is given by

$$W(\boldsymbol{r}) = \int_0^{D/2} d\rho \rho \int_0^{2\pi} d\alpha h(h^2 + (x - \rho \cos\alpha)^2 + (y - \rho \sin\alpha)^2)^{-3/2}/2\pi$$

Accordingly, the corresponding tSOT temperature is given by

$$\delta T_{tSOT}(x,y) = (4\eta/\pi D^2) \int dx' \int dy' W(x - x', y - y')\delta T(x', y'),$$

where $\eta$ is a proportionality constant determined by the heat flow balance conditions. Since in this section we are interested only in the spatial resolution of the tSOT response function in the ballistic regime we take $\eta = 1$ for simplicity. Accordingly, the kernel $W(\boldsymbol{r})$ is properly normalized such that a uniform surface temperature $\delta T(x,y) = \delta T$ results in $\delta T_{tSOT}(x,y) = \delta T$.

In the same vein, for the case of a 1D heat source, where $\delta T(x, y) = \delta T(x)$, we deduce a 1D tSOT kernel function

$$W_1(x) = \int dy\, W(x,y) = (h/\pi) \int_0^{D/2} d\rho \rho \int_0^{2\pi} d\alpha\, (h^2 + (x - \rho \cos\alpha)^2)^{-1}$$

and

$$\delta T_{tSOT}(x) = (4\eta/\pi D^2) \int dx' W_1(x - x')\delta T(x').$$

Figure S3a presents the 2D kernel $W(\boldsymbol{r})$ at different normalized heights $h/D$, indicating a highly localized behavior up to $h/D \cong 0.5$. We define the spatial resolution of the sensor for given tSOT diameter $D$ and scanning height $h$ as the full width at half maximum (FWHM) of $W(x, y = 0)$. Figure S3b plots the resulting spatial resolution as a function of $D$ and $h$ with dashed lines indicating contours of 50, 100 and 150 nm spatial resolution. Note that for small $h$ the contours are essentially vertical, showing that for $h \ll D$ the spatial resolution is equal to $D$. Even at larger heights of $h \cong D/2$ a fair resolution is attained. For example, a tSOT of $D = 75$ nm at $h = 50$ nm provides a spatial resolution of close to 100 nm.

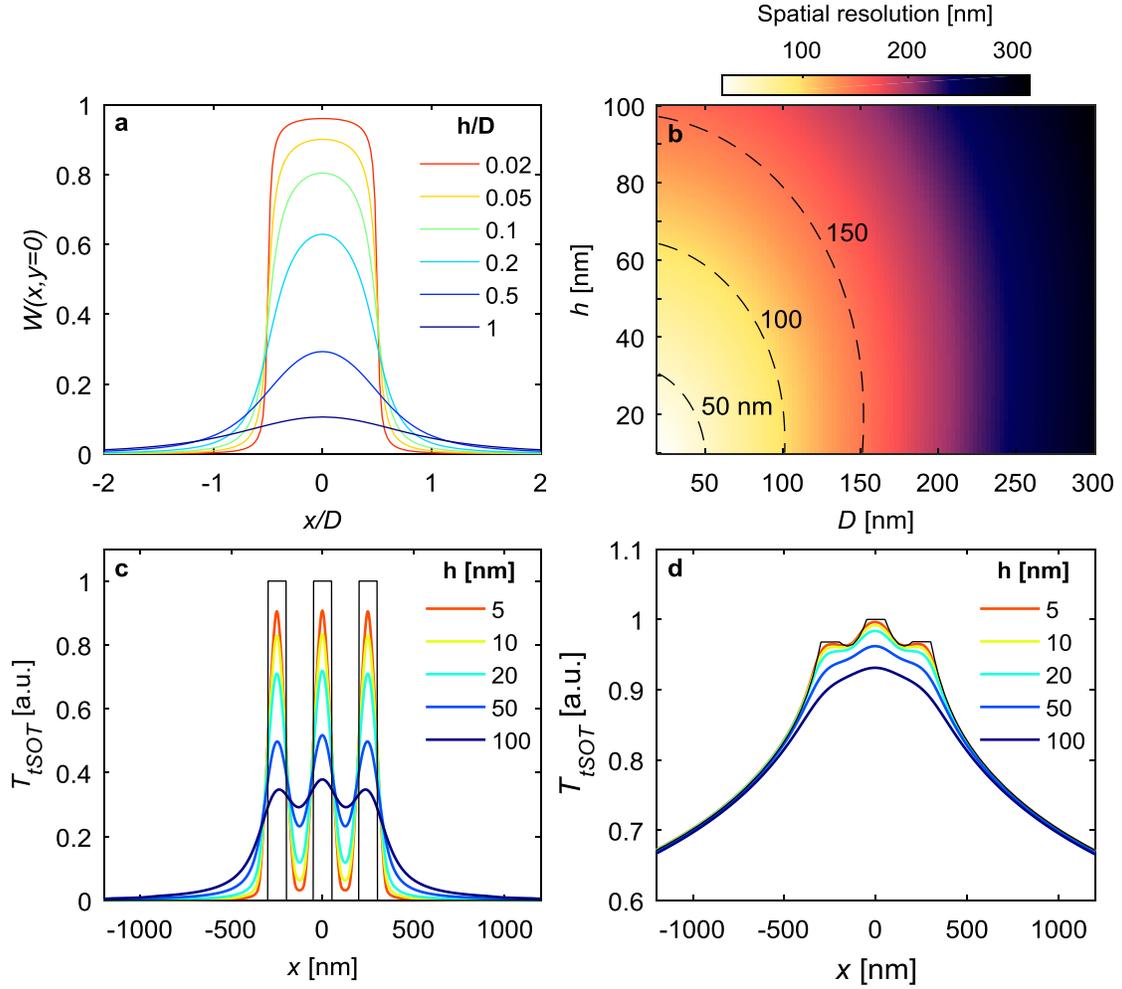

**Figure S3 | Theoretical analysis of the spatial resolution of tSOT. a,** The tSOT 2D heat transfer kernel function for different heights $h$ above a planar sample for ballistic transport of helium atoms. **b,** The resulting spatial resolution of the tSOT technique (defined as FWHM of the kernel function) as a function of tSOT diameter $D$ and scanning height $h$ with contour lines marking resolution of 50, 100, and 150 nm. **c,** Calculated thermal signal measured by tSOT of $D = 100$ nm at several scan heights above a square surface temperature profile $\delta T(x,y)$ with 100 nm wide steps and 150 nm gaps without heat diffusion in the substrate as indicated by black line. **d,** Calculated surface temperature (black line) and the tSOT thermal signal as in **c** in presence of heat diffusion in the substrate. The shape of the thermal profiles is independent of heat conductivity of the substrate.

Figure S3c presents the calculated tSOT signal for a $D = 100$ nm device at different heights above a 100 nm square-shaped surface temperature profile $\delta T(x)$ as indicated by the black line. The tSOT of 100 nm diameter can clearly resolve such temperature profile at low scan heights. However, in practical planar samples, such square-shaped surface temperature profile cannot be readily realized due to heat diffusion in the substrate which causes a significant broadening of the surface temperature. This broadening can be understood by noting that the diffusion equation for a 1D heat source and a homogeneous 3D substrate gives a heat flux that drops as $1/r$, where $r$ is the radial distance from the line-source. This results in a $\log(r)$ temperature decay. Note that this long-range temperature decay is universal and is independent of the thermal conductivity of the substrate.

Figure S4a presents a numerical solution of the heat diffusion equation showing the temperature distribution $\delta T(x,z)$ in the bulk of a semi-infinite sample containing a 1D heater of 110 nm × 30 nm cross section extending in the $y$ direction. The black curve in Fig. S4b shows the corresponding surface temperature $\delta T(x)$ of the sample displaying the long-range decaying tail that is well described by $\log(r)$ dependence (dashed curve). The colored curves in Fig. S4b show the calculated $\delta T_{tSOT}(x)$ profiles for a $D = 100$ nm tSOT at various heights $h$ above the surface. For small $h$ values the tSOT kernel causes rounding of the sharp features on the scale $x < D$ but otherwise $\delta T_{tSOT}(x)$ closely follows the surface $\delta T(x)$. In particular, for $x > \sim 2D$ in the slow $\log(r)$ tail region, the profile $\delta T_{tSOT}(x)$ is almost indistinguishable from the surface $\delta T(x)$ for up to $h \cong 100$ nm. This result demonstrates that at larger distances the measured $\delta T_{tSOT}(x)$ is dominated by heat diffusion in the substrate rather than the intrinsic spatial resolution of the tSOT. Moreover, the $\log(r)$ tail is a characteristic fingerprint of the actual surface temperature of the sample and cannot arise from the tSOT heat transfer kernel.

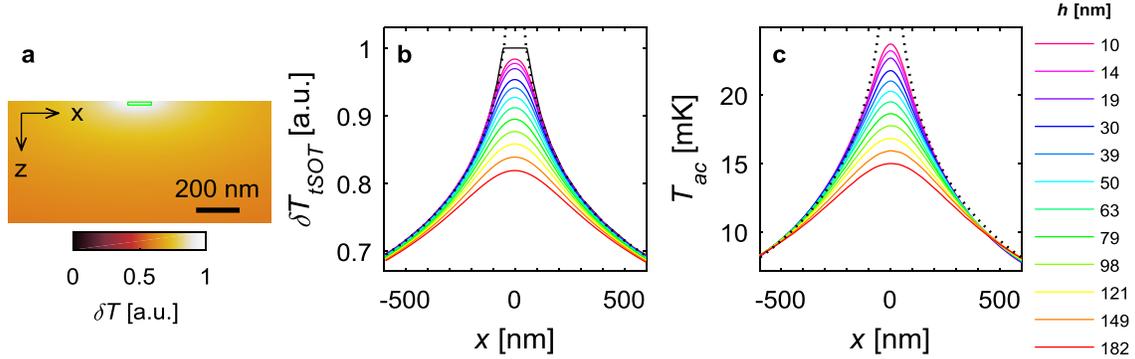

**Figure S4 | Comparison of heat profiles above a 1D heater: simulation vs. experimental results. a,** Numerical solution of the temperature distribution in the bulk of the sample $\delta T(x,z)$ comprising of an infinite slab heater of 110 nm width and 30 nm thickness (outlined in green), embedded in a semi-infinite homogeneous substrate. The heat conductivity within the heater cross section is taken to be much higher than that of the substrate. The resulting functional dependence of the temperature distribution in the bulk is characterized by a universal $\log(r)$ dependence independent of the heat conductivity of the substrate. **b,** The surface temperature of the sample $\delta T(x, z = 0)$ (black line) and the calculated temperature $\delta T_{tSOT}(x)$ of a $D = 100$ nm tSOT scanned at different heights $h$ (as indicated). **c,** The experimentally measured temperature profiles $T_{ac}(x)$ at different scan heights above a nanowire heater. The sample and the tSOT are described in Fig. S5, where the current $I_{ac}$ was applied only to the central nanowire. The dotted lines in **b** and **c** are fits to $\log(r)$ asymptotic behavior, characterizing the long-range heat diffusion in the substrate.

Figure S3d shows a similar calculation for the case where three such 1D heaters are placed 150 nm apart of each other in a semi-infinite substrate. The black curve, which describes the surface $\delta T(x)$, indicates that the heat signals of the three heaters strongly overlap due to the slow $\log(r)$ tail, producing only weak shoulders at the heater positions. We emphasize that the surface $\delta T(x)$ in this case is independent of the value of the heat conductivity of the substrate (up to a constant factor). The corresponding $\delta T_{tSOT}(x)$ well resolves the shoulders up to $h \cong D/2$ while further away from the heaters the profile $\delta T_{tSOT}(x)$ is essentially indistinguishable from $\delta T(x)$ at even higher scan heights, demonstrating once more the dominance of substrate heat diffusion in the thermal signal.

## S4. Experimental demonstration of the tSOT spatial resolution

In order to demonstrate the actual spatial resolution limit of the tSOT sensor, one has to design an experiment in which the surface temperature changes on short length scales and the dominant role of the heat diffusion in the substrate is reduced. For this purpose three parallel Pd-Au nanowire structures were fabricated on $SiO_2/Si$ substrate ($SiO_2$ thickness of 150 nm). Each nanowire is 110 nm wide and 30 nm thick including 5 nm Ti adhesion layer, with 155 nm gap between the nanowires. A trench of 1.4 µm width and 117 nm depth was then etched into the $SiO_2$ layer using buffered HF solution, resulting in a suspended segment of the three nanowires as shown in tilted SEM image in Fig. S5a. The device was thermally imaged in various current configurations with a tSOT of effective magnetic diameter of 55 nm (determined from the period of tSOT quantum interference) and physical inner and outer diameters of 20 nm and 100 nm respectively.

We first drive an *ac* current $I_{ac} = 0.7$ µA at 663 Hz through the central nanowire resulting in an average dissipation of 8 pW per square segment of the wire and measure $T_{ac}$ of the tSOT across the wire at different scanning heights $h$ as shown in Fig. S4c. The experimental data are in very good qualitative agreement with our numerical results in Fig. S4b. In particular, the $\log(r)$ temperature decay (dashed line) is clearly reproduced, with the very weak $h$ dependence away from the origin confirming the dominant role of heat diffusion in the substrate and the locality of the tSOT kernel. We note that the qualitative behavior is not affected significantly by the fact that the nanowire is suspended. This is due to the fact that the suspended segment is short (1.4 µm) relative to the full length of the nanowire and the suspension height is only 117 nm. As a result, the long-range heat diffusion in the substrate is still governed by a 1D heat source, which gives rise to the observed $\log(r)$ behavior. Next, we applied a total of $I_{ac} = 0.5$ µA at 663 Hz to the three nanowires in parallel and imaged the temperature distribution at $h \cong 15$ nm as shown in Fig. S5b. A line cut through the data is shown in Fig. S5c. The presence of the two outer nanowire heaters is visible as shoulders in the $T_{ac}$ profiles followed by $\log(r)$ tails in very good qualitative agreement with the numerical results in Fig. S3d. Note that even though $T_{ac}$ in the suspended segment in Fig. S5b (outlined by the dotted lines) is higher than in the non-suspended regions, the $\log(r)$ decay through the substrate is essentially the same as discussed above.

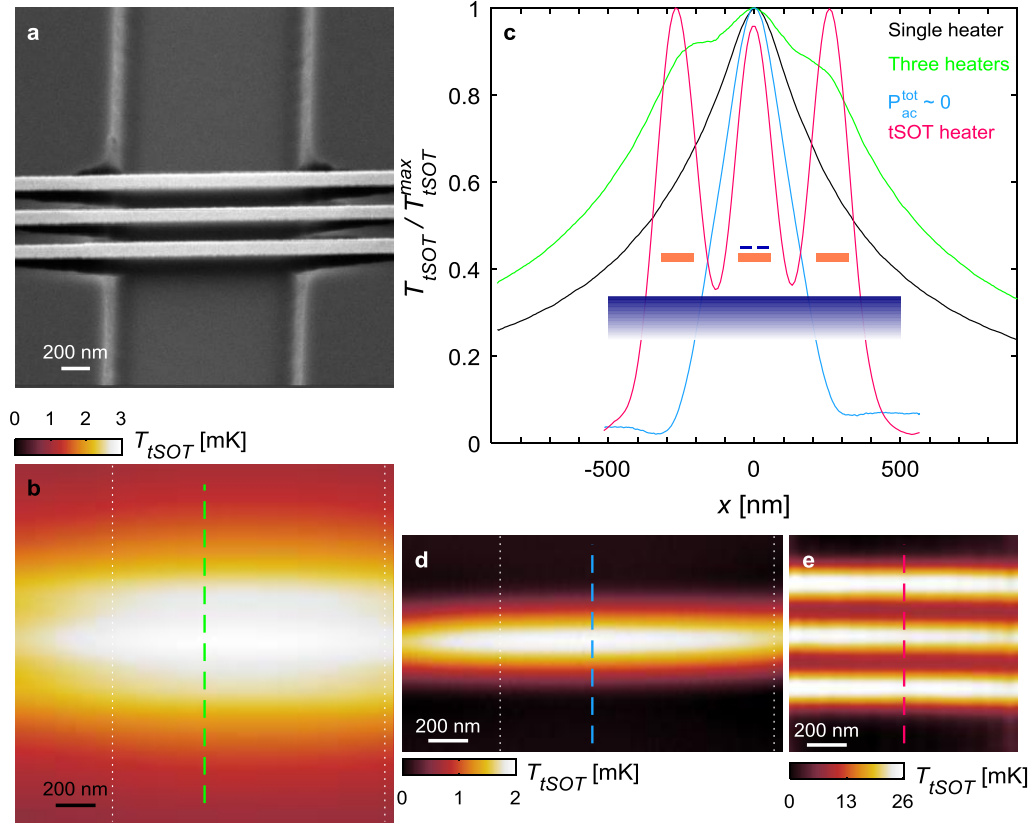

**Figure S5 | Experimental demonstration of spatial resolution of tSOT. a,** SEM image (tilted at 45°) of three Pd-Au nanowire heaters (110 nm wide, 30 nm thick, 155 nm separation) suspended above a 1.4 µm wide and 117 nm deep trench etched into SiO$_2$/Si substrate. **b-d,** Thermal imaging $T_{tSOT}(x,y)$ of the device in three different measurement schemes using tSOT of $D = 100$ nm at scan height of $h = 15$ nm. **b,** Thermal image due to *ac* current applied in parallel to all three nanowires dissipating 3 pW per square segment of nanowire measured at the second harmonic of 663 Hz. White dotted lines indicate edges of the trench. **c,** Thermal profiles across the device for the cases of central wire heater, three wire heaters (along green line in **b**), compensated total power (along blue line in **d**), and tSOT heating (along red line in **e**). Inset: a schematic to-scale cross section of the experimental setup including the substrate (light blue), nanowires (orange), and tSOT apex (blue). **d,** Thermal image for the case of "heat dipole" in which $P_{ac} \cong 8.7$ pW per square segment of nanowire is dissipated in the central wire while out-of-phase power of $P_{ac} \cong -4.2$ pW and -4.5 pW per square segment of nanowire is "absorbed" in the two outer wires thus suppressing the long-range heat diffusion in the substrate (all measured at the second harmonic of 663 Hz). **e,** Thermal image with no current flowing in the wires while a larger current of 152 µA is driven through the tSOT causing dissipation in the tip apex. The resulting increase in the local temperature in the suspended nanowires during scanning is larger than in the recessed substrate as expected from the ballistic model. The scan window is within the suspended section. Time constants are 30 ms/pixel in all images, pixel sizes are 13 nm (b), 10 nm (d) and 10 nm (e).

We emphasize that the temperature $T_{tSOT}(x,y)$ measured by the non-contact tSOT scanning at height $h$ above the sample reflects the local temperature profile of the sample convoluted with the height dependent kernel $W$. Consequently, in the case of suspended nanowires the $T_{tSOT}(x,y)$ reflects the temperature of the nanowires as well as the background temperature of the substrate in-between the wires. In contrast, contact-based vacuum SThM[11-15] would detect only the temperature of the suspended nanowires if scanning at a height corresponding to the top surface plane of the nanowires.

The above cases indicate that heat diffusion in the bulk has a dominant effect on the surface temperature distribution. This long-range contribution, however, can be significantly suppressed by two methods as follows. The first approach takes advantage of the fact that the phase of the *ac* power $P_{ac}$ can be readily controlled. Consider a wire carrying a low frequency *ac* current $I\cos(\omega t)$, which results in power dissipation $P = P_0 \cos^2(\omega t) = 0.5 P_0 + 0.5 P_0 \cos(2\omega t) = P_{dc} + P_{ac}$. As a result, the wire creates a certain surface temperature profile $\delta T(x,t) = \tau(x) + \tau(x)\cos(2\omega t)$. Note that the temperature profile for the same wire carrying current of $I\sin(\omega t)$ will be $\delta T(x,t) = \tau(x) - \tau(x)\cos(2\omega t)$ with a negative *ac* term. Let us consider two such wires at $x = -d$ and $x = d$ carrying currents $I\cos(\omega t)$ and $I\sin(\omega t)$. The resulting temperature distribution will now be given by $\delta T(x,t) = \delta T_{dc}(x) + T_{ac}(x)\cos(2\omega t) = \big(\tau(x-d) + \tau(x+d)\big) + \big(\tau(x-d) - \tau(x+d)\big)\cos(2\omega t)$. Since the instantaneous $P$ is always positive, the instantaneous $\delta T(x,t)$ and $\delta T_{dc}$ are positive as well. However, $T_{ac}(x)$ can be either positive or negative. In particular, when the two wires overlap ($d = 0$), $T_{ac}(x) = 0$ everywhere and $\delta T(x,t) = 2\tau(x)$. This is because $P_{ac}$ of the first wire is fully "absorbed" by the negative $P_{ac}$ of the second wire. When the two wires are separated, the total $P_{ac}^{tot}$ remains zero and the corresponding $T_{ac}(x)$ attains a "dipole" structure with positive and negative values on the two wires. As a result, the long-range $\log(r)$ decay of $T_{ac}(x)$ of a single wire is replaced by a faster $1/r$ behavior. Applying this additional degree of freedom to three wires we can produce much more local $T_{ac}(x)$ temperature profiles. We implement this concept here by driving $I_{ac}$ in the central wire and $\sim I_{ac}/\sqrt{2}$ with a phase shift of $\pi/2$ in each of the outer wires. As a result, the power $P_{ac}$ dissipated in the central wire is mostly "absorbed" by $-P_{ac}/2$ in each of the outer wires thus forming a "screened" local source with total $P_{ac}^{tot} \sim 0$. The resulting $T_{ac}$ image in Fig. S5d and the corresponding line profile in Fig. S5c show a narrow temperature peak and the strong suppression of the $\log(r)$ tail, demonstrating the high spatial resolution of the tSOT. The above three current schemes are summarized by numerical simulations in Fig. S6. The temperature distribution in the bulk of the substrate for the cases of single heater, three heaters, and $P_{ac}^{tot} \sim 0$ are shown in Figs. S6a-c, and the corresponding $\delta T_{tSOT}(x)$ for $D = 100$ nm at $h = 15$ nm are presented in Fig. S6d demonstrating a proper description of the experimental data. In particular, the width of the measured $T_{tSOT}$ signal at FWHM of the $P_{ac}^{tot} \sim 0$ of 284 nm in Fig. S5c is in excellent agreement with the calculated width of 270 nm in Fig. S6d.

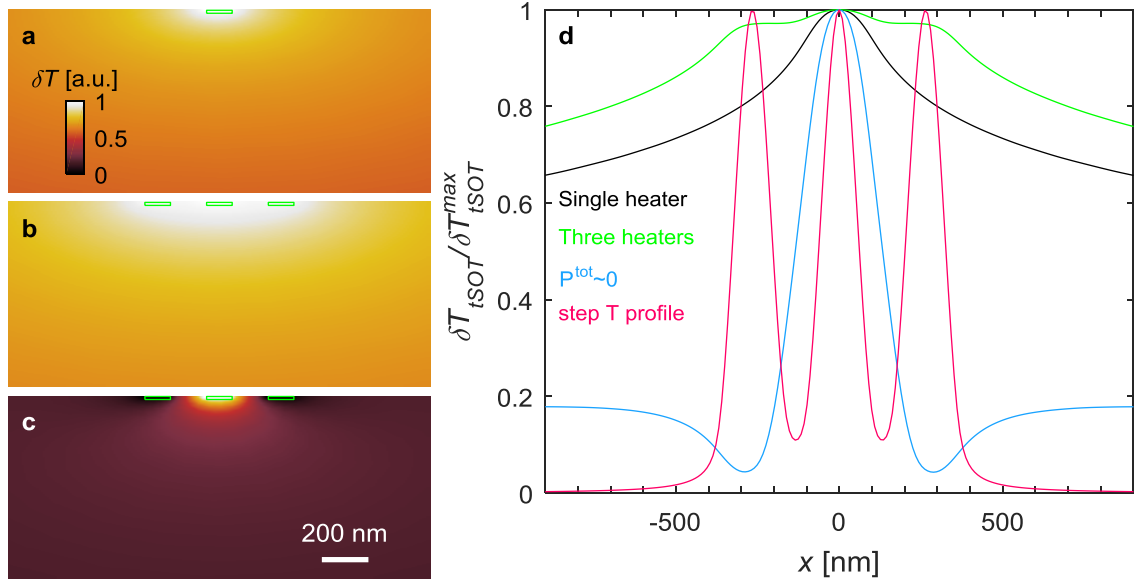

**Figure S6 | Theoretical comparison of different heating schemes. a,** Temperature distribution $\delta T(x,z)$ in the bulk of the sample due to a 1D heater of 110 nm × 30 nm cross section (green outline) embedded in a semi-infinite substrate. **b,** Temperature distribution due to three such heaters separated by a gap of 155 nm. **c,** Temperature distribution in the case of a "heat dipole" with $P_{ac}^{tot} \sim 0$ in which each of the two outer heaters "absorbs" 43% of the *ac* power dissipated by the central heater (see text). **d,** Calculated temperature profiles for $D = 100$ nm tSOT scanning at $h = 15$ nm above the surface in the three cases **a**-**c** and for the case of local heating by the tSOT.

Finally, we present an alternative approach to reduce the long-range substrate heat diffusion contribution based on the fact that for the case of a point source the heat flux in the substrate drops as $1/r^2$ resulting in $1/r$ decay of the surface temperature rather than the $\log(r)$ behavior for 1D heat source. In order to implement a point source we can use the tSOT itself as a point heater by driving a current through the tip, significantly larger than the critical current, thus causing a local heating of the tSOT apex. This point-like heater will cause local heating of the sample surface which in turn causes a corresponding change in the tSOT temperature. A similar mechanism is commonly used in SThM schemes in which a warm scanning tip acts as a local heat source[40,41]. In our sample geometry and in the ballistic He transport regime, the tip that scans at a constant height will locally heat up the suspended nanowires when hovering 15 nm directly above their surface. When passing above the gaps between the wires, in contrast, the local heating of the etched $SiO_2$ substrate that resides some 162 nm below the tip will be significantly smaller due to the much larger solid angle, leading to a sharp temperature contrast between the nanowires and the substrate. Figure S5e presents a thermal image of the suspended section in this "tSOT heater" mode showing three sharply resolved nanowires. The line cut through the image is presented in Fig. S5c with a width of the central peak at FWHM of only 132 nm as compared to 110 nm width of the nanowire and the tSOT diameter of 100 nm. We can model this situation by approximating a step-like change in the effective surface temperature of a sample as shown in Fig. S3c. In this case the rounding of the thermal signal will be determined only by the tSOT kernel. Figure S6d "step T profile" presents the calculated $\delta T_{tSOT}(x)$ profile using $h = 15$ nm and $D = 100$ nm which shows a very good correspondence with the experimental data albeit with a larger temperature modulation. Since the width of the nanowires is comparable to $D$ the actual effective surface temperature of the sample due to the heating by the scanning tSOT is expected to be also significantly rounded explaining the reduced modulation.

Interestingly, in Fig. S5e a higher $T_{tSOT}$ temperature is attained above the suspended nanowires in contrast to what one may expect. This counter-intuitive behavior results from the height differences and the fact that the He atoms are in ballistic rather than diffusive regime. The local density of heat flow from the warm tip to the sample and the corresponding local surface temperature increase are strongly height dependent due to solid angle considerations. Since the suspended nanowires are much closer to the tip, their local temperature increase is significantly larger as compared to the temperature increase of the deeper Si substrate between the wires, leading to the increased $T_{tSOT}$ above the nanowires. Indeed, similar measurements in the planar region where the nanowires reside on the substrate show a lower $T_{tSOT}$ above the nanowires due to their higher heat conductivity.

The above experimental results in various dissipation schemes are well described by our numerical calculations and clearly demonstrate that the spatial resolution of the tSOT at low scanning heights is comparable to the tip diameter $D$.

## S5. Demonstration of tSOT temperature sensitivity

Our various tSOT sensors display thermal sensitivities of the order of 1 µK/Hz$^{1/2}$. Figure S7a demonstrates imaging of such low thermal signals in practice, using a 120 nm wide Cu nanowire on $SiO_2$/Si substrate ($SiO_2$ thickness of 1 µm) carrying an *ac* current causing an *ac* heat dissipation in the nanowire at $2f = 13.1$ kHz. Figure S7 presents two $T_{ac}$ profiles by scanning with the 46 nm diameter tSOT (same tSOT as used for Fig. 1) across the nanowire at two levels of power dissipation $P$ per square segment of the nanowire. The blue data were acquired at $P = 143$ pW while the red data were measured at $P = 57$ fW close to the Landauer limit of 40 fW for 1 GHz qubit operation. The $T_{ac}$ profiles show a sharp peak above the nanowire and extended tails due to heat diffusion in the $SiO_2$/Si substrate.

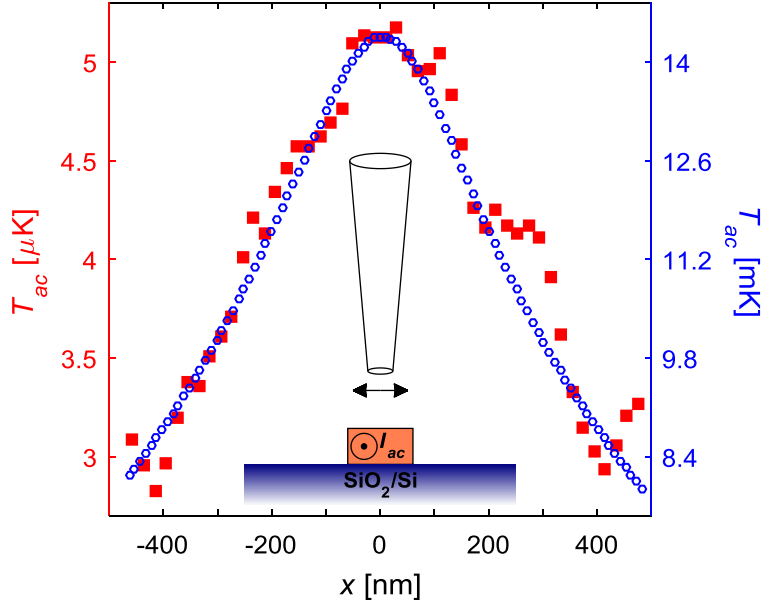

**Figure S7 | Demonstration of tSOT temperature sensitivity.** $T_{ac}$ profiles across a 120 nm wide Cu nanowire at dissipation powers of 143 pW (blue) and 57 fW (red) per square segment measured by a tSOT of 46 nm diameter. The data were acquired using lock-in amplifier time constants of 0.3 sec and 1 sec respectively at height of 120 nm above the surface. Inset: schematic diagram of the Cu nanowire sample and the scanning tSOT.

## S6. Assessment of thermal resistances of the tSOT

Figure 1d of the main text shows a simplified low-frequency thermal model of the tSOT scanning system. Evaluation of $R_{sd}$ and $R_{ss}$ is thus highly desirable for proper utilization and optimization of the tSOT sensor. In the following we assess theoretically the sensor-device thermal resistance $R_{sd}$ and then determine experimentally the corresponding sensor-support resistance $R_{ss}$.

Since at 4.2 K and 1 mbar pressure the mean free path of He atoms is ~3 μm we consider for our scanning heights a molecular flow regime in which He atoms flow ballistically between the sample surface and the apex of the tSOT sensor. According to Kennard's law[39] the net power flux $\Omega$ per unit temperature hitting a unit area under this regime is $\Omega = \left( \alpha \frac{\gamma+1}{\gamma-1} \sqrt{\frac{R}{8\pi MT}} \right) p \equiv \beta p$. Here $\alpha$ is the accommodation coefficient (here taken as unity), $\gamma$ is the adiabatic constant, $R$ is the universal gas constant, $M$ is the molar mass, $p$ the gas pressure, and $T = 4.2$ K the surface temperature (assumed to be much larger than the induced *ac* temperature variations) resulting in $\beta = 18$ m/s/K. $R_{sd}$ is then given by $R_{sd} = 1/(\beta S p)$, where $S = \pi D^2/4$ is the surface area of a tSOT of diameter $D$. Figure S8 shows the resulting $R_{sd}(p)$ calculated for $D = 240$ nm tSOT vs. the He pressure $p$. $R_{sd}$ can be readily varied from about $1.6 \times 10^{10}$ K/W at 1 mbar to $5 \times 10^{8}$ K/W at 30 mbar.

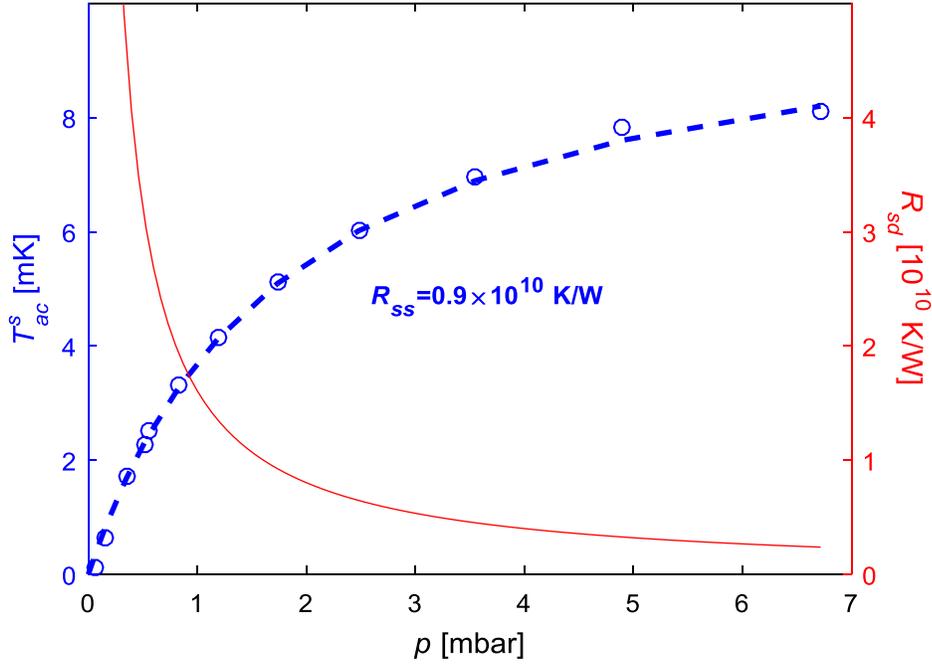

**Figure S8 | Thermal resistances of tSOT.** Red: Calculated thermal resistance $R_{sd}$ between the sensor and the sample device for 240 nm tSOT vs. the He pressure $p$ in the molecular flow regime. Blue circles: measured tSOT *ac* temperature $T_{ac}^s$ at height of 100 nm above a thin film heater vs. He pressure. Dashed blue line is a theoretical fit resulting in thermal resistance between the tSOT sensor and its quartz support structure of $R_{ss} = 0.9 \times 10^{10}$ K/W.

Solving the effective circuit of Fig. 1d, the relation between the *ac* temperatures $T_{ac}^s$ of the tSOT and $T_{ac}^d$ of the device is given by $T_{ac}^s/T_{ac}^d = [1 + R_{sd}/R_{ss}]^{-1} = [1 + (R_{ss}\beta Sp)^{-1}]^{-1}$. A tSOT sensor of 240 nm diameter was held at a height of 100 nm above the surface of a metallic thin film heater of 170 Ω resistance driven by an *ac* current of 10 μA at 3.7 kHz and the $T_{ac}^s$ was measured as a function of the He pressure $p$ as shown in Fig. S8 (blue circles). The dashed line shows an excellent fit to the data using the above equation with resulting $T_{ac}^d = 10.4$ mK and $R_{ss} = 0.9 \times 10^{10}$ K/W. Such remarkably high thermal resistance between the sensor and the supporting quartz pipette, even for the extreme case of a large tSOT (relative to other tSOT devices used throughout the presented work), cannot be explained by naively taking into account the bulk thermal conductivity of quartz. The nanoscale cross section of the pipette, however, quenches the radial phonon degrees of freedom resulting in quantum-limited thermal conductance[24,42,43]. A 1D channel with a single phonon mode has a quantum thermal resistance of $R_0 = 3h/(\pi^2 k^2 T) = 2.52 \times 10^{11}$ K/W at 4.2 K. The derived $R_{ss}$ thus implies about 28 quantum channels of thermal conductance in our 240 nm diameter tSOT. The 46 nm tSOT in the main text may thus support only one or two channels resulting in $R_{ss}$ in excess of $10^{11}$ K/W.

An additional insight can be drawn from the presented pressure dependence of the thermal signal regarding the governing heat transfer processes. Since the signal vanishes in the low pressure limit, the possible contribution of near-field radiative processes to the heat transfer between the sample and the tSOT is apparently negligible under the presented working conditions.

## S7. Bandwidth of the tSOT thermal imaging

Thermal processes at the nano-scale may have very short relaxation times and hence it is highly desirable for a nano-thermometer to have a high bandwidth (BW) to study relaxation processes. Figure S9 shows the frequency response of tSOT positioned at a relatively large distance of 500 nm above an Au thin film heater in He exchange gas pressure of 26 mbar. The figure shows the measured $T_{ac}$ vs. the frequency $f$ of the heat signal (second harmonic) for two values of the *ac* power in the heater demonstrating BW of over 200 kHz which was limited by the BW of our electronic readout circuit.

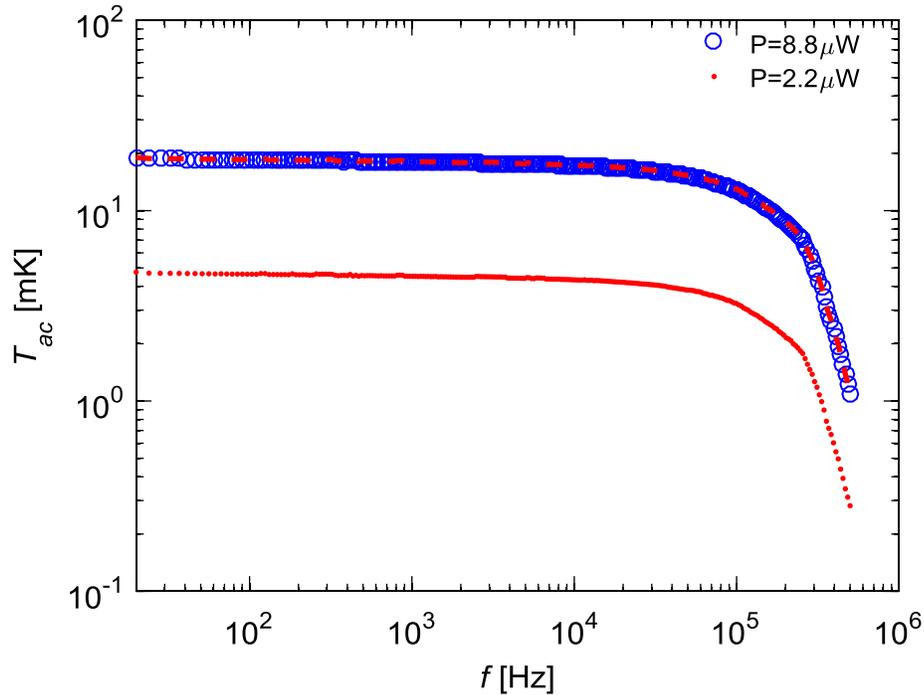

**Figure S9 | Bandwidth characterization of the tSOT.** Frequency dependence of $T_{ac}$ measured by a tSOT 500 nm above Au heater for *ac* heater powers of 8.8 μW and 2.2 μW. Dashed red line shows the data collapse by multiplying the low power signal by 4. The bandwidth of 200 kHz is limited by external readout electronics.

We can estimate the intrinsic thermal BW of the tSOT as following. The thermal time constant of the tSOT driven by the heat source in the sample should be given by sensor thermal mass times the sensor-heater thermal resistance $R_{sd}$. The mass of the Pb apex of a tSOT with diameter D = 100 nm and thickness of 10 nm is approximately $10^{-18}$ Kg and therefore has a heat capacity of about C = $10^{-18}$ J/K, taking Pb specific heat[44] as 0.042 cal/Kg·mole. For $R_{sd}$ that is typically below $10^{10}$ K/W, the resulting intrinsic tSOT bandwidth is therefore above 10 MHz.

## S8. Tuning-fork based height control

In some of the measurements performed throughout this work (graphene measurements and the tSOT spatial resolution measurement) scan heights above the sample were controlled by mechanically attaching the tSOT to a quartz tuning-fork (TF). The used TF is a quartz oscillator with nominal frequency of 32.768 kHz (product number TB38 of HM international). The TF was filed along three planes to adapt it to scanning constraints without violating the inversion symmetry of its two prongs. The filing shifted the resonant frequency to about 37 kHz. The tSOT protruded a few tens of μm above the highest point of the TF as demonstrated in Fig. S10. The TF was mechanically excited using a piezo-electric crystal as a dither. The TF's response was read electrically through the voltage on its electrodes, which was pre-amplified at room temperature and fed into a lock-in amplifier. This assembly has a quality factor of ~30,000 at a temperature of 4.2 K and vacuum conditions. As the tSOT approaches heights of ~10 nm above the surface of the sample its electrostatic interaction with the surface affects the resonant frequency of the TF. By applying a phase-locked loop (PLL) this frequency shift is monitored and used to determine the proximity of the tip to the sample. The scanning is then performed at a constant z-plane at a determined height above the sample surface.

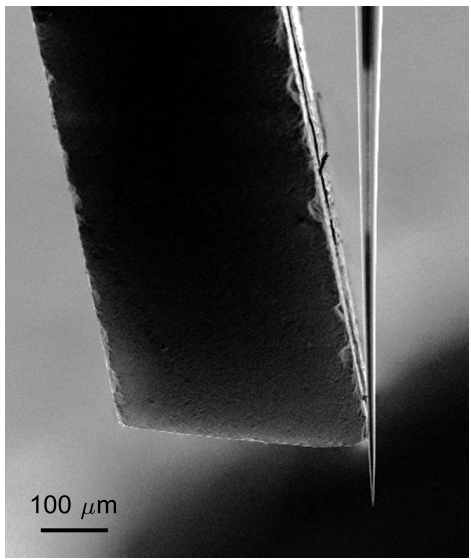

**Figure S10 | Quartz crystal tuning fork resonator for tSOT height control.** A SEM image showing a side view of the filed TF with the attached tSOT pipette. The tSOT protrudes typically 50 to 100 μm beyond the TF.

## S9. Simultaneous magnetic and thermal imaging

By applying a proper combination of magnetic field $H_z$ and $I_{bias}$, the tSOT can be tuned to have both magnetic field and thermal sensitivity, thus providing a unique capability of simultaneous magnetic and thermal imaging. Figure S11a shows a SEM micrograph of a "zigzag" nanostructure consisting of alternating segments of 100 nm wide Cu and 80 nm wide ferromagnetic Py ($Ni_{80}Fe_{20}$) wires. An *ac* current $I_{ac} = 10$ µA is applied to the sample at $f = 6.56$ kHz and three signals are acquired simultaneously from the tSOT during scanning: time-averaged *dc* and two *ac* signals at frequencies $f$ and $2f$ using two lock-in amplifiers. At the used bias conditions the tSOT had a thermal response of -8.6 µA/K and a simultaneous magnetic response of 91 nA/mT.

The *dc* signal provides an image of the *dc* magnetic field $B_z^{dc}$ (Fig. S11b) in presence of out-of-plane $\mu_0 H_z = 0.32$ T and an in-plane field $\mu_0 H_{||} = 0.1$ T applied parallel to the Py nanowires. The image reveals that the Py nanowires are in a single in-plane magnetization domain state giving rise to the sharp peaks in the stray field $B_z^{dc}$ at the ends of the nanowires. The Py nanowires also show a weaker bright signal along their length due to the out-of-plane tilting of the magnetic moments by $H_z$.

Figure S11c presents the first-harmonic *ac* component, which describes the *ac* magnetic signal $B_z^{ac}$, generated by $I_{ac}$. It shows that the current flows continuously along the zigzag structure avoiding the upper protruding parts of the nanostructures. This is quantitatively confirmed by the Biot-Savart law numerical simulation of $B_z^{ac}$ in the inset, also providing an accurate estimate of the tSOT scanning height of 175 nm.

The power dissipated in a resistor $R$ due to a sinusoidal *ac* current $I_{ac} = I_0 \cos \omega t$ is given by $P = RI_{ac}^2 = RI_0^2(1 + \cos 2\omega t)/2$ giving rise to *dc* temperature increase $T_{dc}$ and *ac* temperature variation $T_{ac}$ at $2\omega$. Figure S11d shows the resulting temperature map $T_{ac}$ (the accompanying *dc* signal due to $T_{dc}$ is substantially smaller than $B_z^{dc}$ and was subtracted from Fig. S11b for clarity). In contrast to the very uniform $B_z^{ac}$ along the zigzag structure in Fig. S11c, $T_{ac}$ in Fig. S11d is highly non-uniform revealing a significantly higher temperature of the Py segments. This is the result of about 40 times higher sheet resistance 18 Ω/□ in Py than in Cu nanowires, giving rise to significantly higher dissipation along the Py segments. Interestingly, the nanowire segments protruding beyond the top corners of the zigzag also display an increased temperature despite the fact that no current flows there, revealing higher thermal conductivity of the metallic nanowires as compared to the underlying substrate. The tSOT thus provides multifunctional abilities for simultaneous imaging of magnetic and thermal signals.

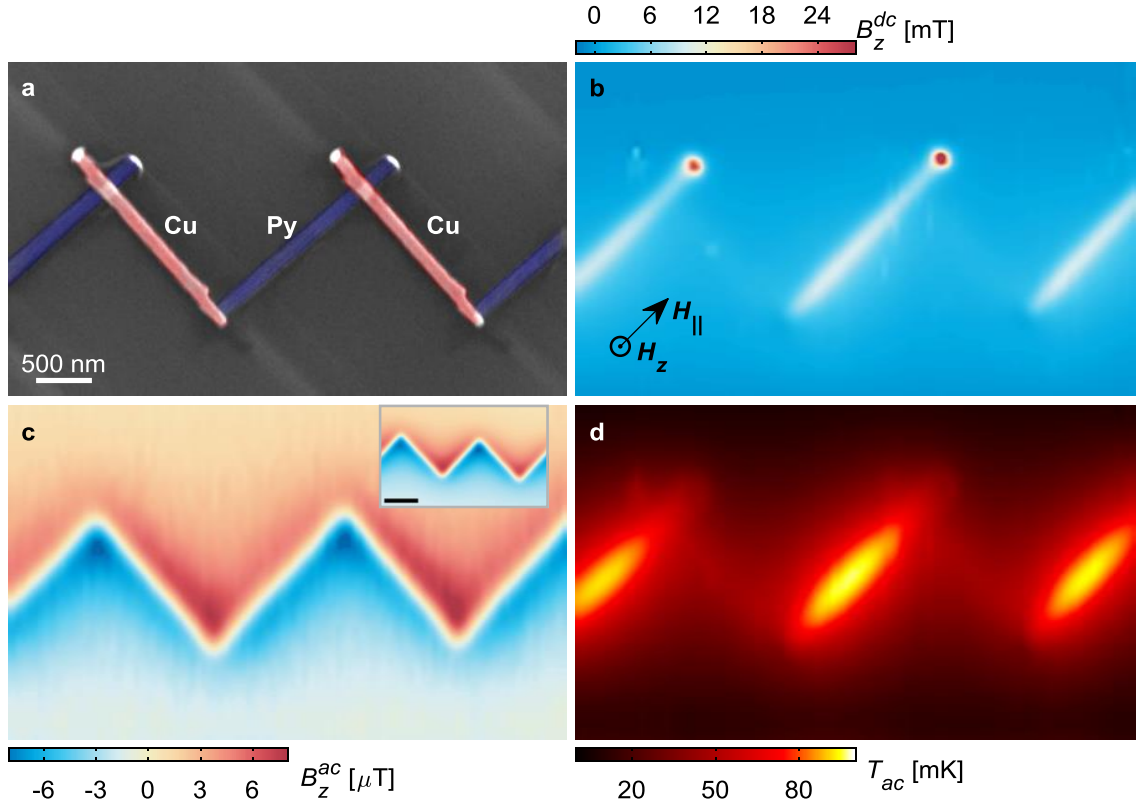

**Figure S11 | Simultaneous scanning tSOT thermal and magnetic imaging. a**, False color SEM image of zigzag structure of alternating Cu (red) and Py (blue) nanowires. **b**, Image of $B_z^{dc}$ in presence of applied magnetic fields of $\mu_0 H_z = 0.32$ T and $\mu_0 H_\parallel = 0.1$ T revealing the in-plane magnetization of Py segments. **c**, $B_z^{ac}$ due to applied $I_{ac} = 10$ μA at 6.56 kHz through the structure. Inset: Theoretically calculated $B_z^{ac}$ 175 nm above the sample. **d**, Simultaneously measured thermal image of $T_{ac}$ at 2*f* showing significantly higher dissipation in Py segments (pixel size 50 nm, scan speed 120 ms/pixel).

## S10. Carbon nanotube fabrication

The SWCNT devices were fabricated on p-doped Si (100) with a thermally grown oxide layer of 300 nm, as reported previously[25]. Briefly, parallel stripes (25 μm wide, 25 nm thick) of amorphous $SiO_2$ coated by a thin layer (nominally 0.3 nm) of Fe growth catalyst were deposited on the substrate by standard photolithography and electron-beam evaporation. The samples were introduced into a tube furnace and were heated to 550° C for 20 min in air to remove organic contaminations. The SWCNT were grown by CVD: samples were heated to 900° C in an atmosphere of 300 sccm Ar and 200 sccm $H_2$, followed by introduction of 1 sccm $C_2H_4$. Growth time was 45 min, by the end of which samples were left to cool in Ar. Electrodes for transport measurements were subsequently patterned by electron beam lithography, followed by deposition of 20 nm of Pd for Ohmic contacts and 10 nm of Au.

## S11. Transport characterization of the CNT devices

Figure S12 presents the electrical characterization of the two SWCNT devices discussed in the main text as a function of the applied back gate voltage $V_g$. Both devices are highly disordered, forming a series of quantum dots, resulting in very low conductance and large fluctuations of the conductance vs. $V_g$. Device 1 (Figs. S12a and 2a) shows a metallic behavior with finite conductance at all values of $V_g$, while device 2 (Figs. S12b and 2b) shows a semiconducting behavior.

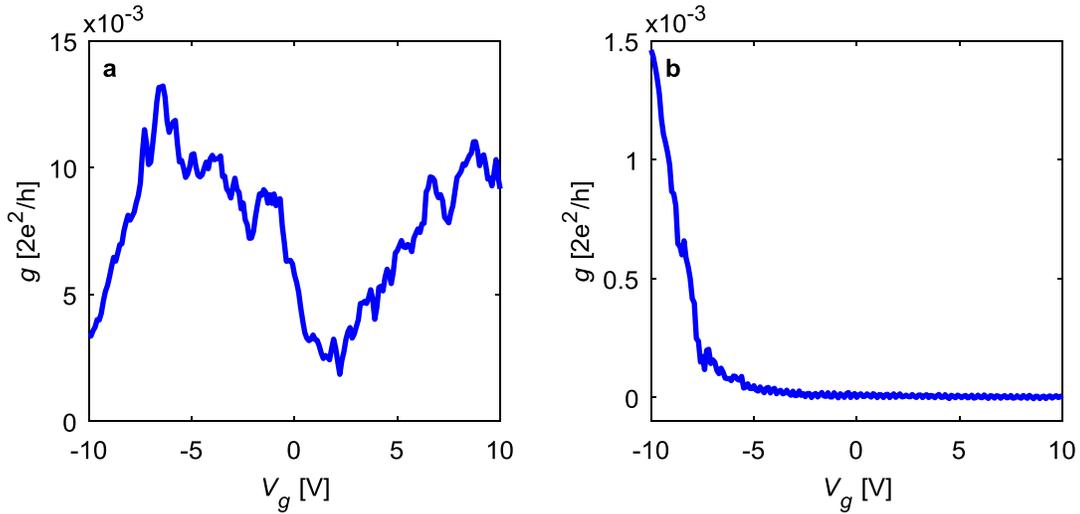

**Figure S12 | Gate dependence of SWCNT conductance at 4.2 K. a**, Device 1 measured with applied *ac* voltage of 80 mV at 31.5 Hz showing a highly disordered metallic behavior. **b**, Device 2 measured with *ac* voltage of 110 mV at 28.9 Hz showing a semiconducting behavior.

## S12. hBN encapsulated graphene device fabrication

The devices were fabricated from single-layer graphene encapsulated between crystals of hexagonal boron nitride (hBN) (40 nm thick bottom crystal and 12 nm top). The transfer of the crystals was carried out using the dry-peel technique described previously[45]. The heterostructures were assembled on top of an oxidized Si wafer (90 nm of SiO$_2$) which served as a back gate, and then annealed at 300 C in Ar-H$_2$ atmosphere for 3 hours. A PMMA mask defining the shape of the device was fabricated on top of the annealed hBN-graphene-hBN stack by electron-beam lithography. The stack was etched in O$_2$/CHF$_3$ plasma with fast selective removal of hBN[46]. Metallic contacts (Niobium) were implemented in a second stage where an additional PMMA mask was used for both etching the stack and for the metal lift-off[46]. This procedure is similar to previous reports and results in electrons mobility above 30 $m^2/V \cdot s$ and ballistic transport on length scale of many microns[28,47].

## S13. Ring-like patterns in scanning gate thermometry due to localized states

In order to provide qualitative description of the characteristic ring-like patterns in scanning gate thermometry as seen in Figs. 2, 3, S15, and S16 we performed 3D heat diffusion simulations using COMSOL. The CNT was modeled as a nanowire in a substrate with looped form matching the actual shape in Fig. 2d. Due to numerical limitations the CNT was modeled with a cross section of 50×50 nm$^2$ and the heat conductivity of the CNT (1100 W·m$^{-1}$K$^{-1}$ for 5 nm diameter CNT [48]) was rescaled accordingly to give an equivalent heat flow along the CNT. Similarly, the graphene was modelled as 50 nm thick with rescaled heat conductivity equivalent to 2000 W·m$^{-1}$K$^{-1}$ for thickness of 0.3 nm [48] with geometry as in Fig. 3a. The heat conductivity of semi-infinite SiO$_2$ substrate was taken as 0.08 W·m$^{-1}$K$^{-1}$ [49]. The qualitative features presented below are insensitive to the values of thermal conductivities, which mainly determine the overall temperature scale.

Figure S13a shows the surface temperature distribution due to dissipation per unit length in the CNT of 20 pW/μm showing the enhanced temperature along the CNT and the heat diffusion in the substrate on a typical length scale of 1 μm, similar to the experimental data in Fig. 2a (in the simulation straight segments of dissipating CNT extend 7 μm on each side beyond the displayed area). Figure S13b shows the surface temperature distribution caused by a localized resonant QD state on the CNT (located as marked by the right blue dot in Fig. S13a), modeled as a region with 20 nm diameter dissipating 20 pW giving rise to a heat spot diffusing into the substrate. The visible anisotropic shape of the surface temperature is due to higher heat conductivity along the CNT. In the case of reduced heat conductivity along the CNT the heat spot will be isotropic with universal $1/r$ decaying temperature profile determined by heat diffusion in an isotropic substrate. Temperature distribution due to another independent QD is shown in Fig. S13c.

When the QDs are not in resonance the thermal image acquired by the scanning tSOT will be described by Fig. S13a. On the other hand, if one of the QDs is at resonance the excess dissipation will result in a thermal image that is given by the sum of Figs. S13a and S13b or of Figs. S13a and S13c accordingly. However, the excess dissipation is present only when the QD is tuned into one of its Coulomb blockade resonances $n$ by the tip potential. This occurs when the tSOT is at some lateral distances $R_n$ from the QD (which depends among other things on the scanning height and the tip potential $V_{tSOT}$). As a result, the excess dissipation in the QD is present only when the tSOT is located along concentric rings of radius $R_n$ shown by the green circles in Figs. S13b,c (only a pair of $R_n$ rings is shown for each QD for clarity). Each QD will have a different set of $R_n$ depending on the local potential structure, doping, and the QD capacitance as observed by scanning gate AFM[26,50-53]. The width of the rings is determined by the broadening of the Coulomb blockade energy levels of the QD and was taken to be 50 nm in the simulations.

The thermal image attained by the scanning tSOT will therefore be composed as following. At any location except at the rings, the image will be given by Fig. S13a. When the tip is on one of the rings of the right QD, however, the QD dissipates and hence the image will be given by the sum of Figs. S13a and S13b. Similarly, when the tSOT is scanned across the left QD, the measured temperature along the green circles will be given by addition of Figs. S13a and S13c. The resulting combined thermal image is presented in Fig. S13d providing a qualitative description of the experimental results in Figs. 2 and S15. The same process is presented in Figs. S13e-h for the case of the graphene washer on $SiO_2$ substrate, where a heat of 40 pW/μm$^2$ was dissipated uniformly in the graphene. The resulting surface temperature of the sample is presented in Fig. S13e showing a uniform temperature in the graphene and a decaying distribution outside the washer and in the central aperture due to heat diffusing in the substrate, similar to the experimental data in Figs. 3b and S16. Figures S13f and S13g present the thermal map due to 60 pW dissipation at localized resonant states on dots with a diameter of 20 nm at locations indicated by blue dots in Fig. S13e. Similarly to the case of CNT the excess dissipation at the localized states occurs only when the scanning tip is located at lateral distance $R$ marked by the green circles in Figs. S13f-g. The resulting thermal image attained by a scanning tSOT will therefore appear as shown in Fig. S13h. As discussed in section S16, we attribute the localized states to vacancies and adatoms at the graphene edges, which result in localized states with narrow energy levels pinned near the Dirac point[31]. As a result, in contrast to the multiple energy levels in QD in CNT, we expect only one resonant state in graphene defects, giving rise to a single ring around each defect. The radius of the ring will vary upon changing the tip potential $V_{tSOT}$ as observed experimentally in Fig. S16 and Video S1.

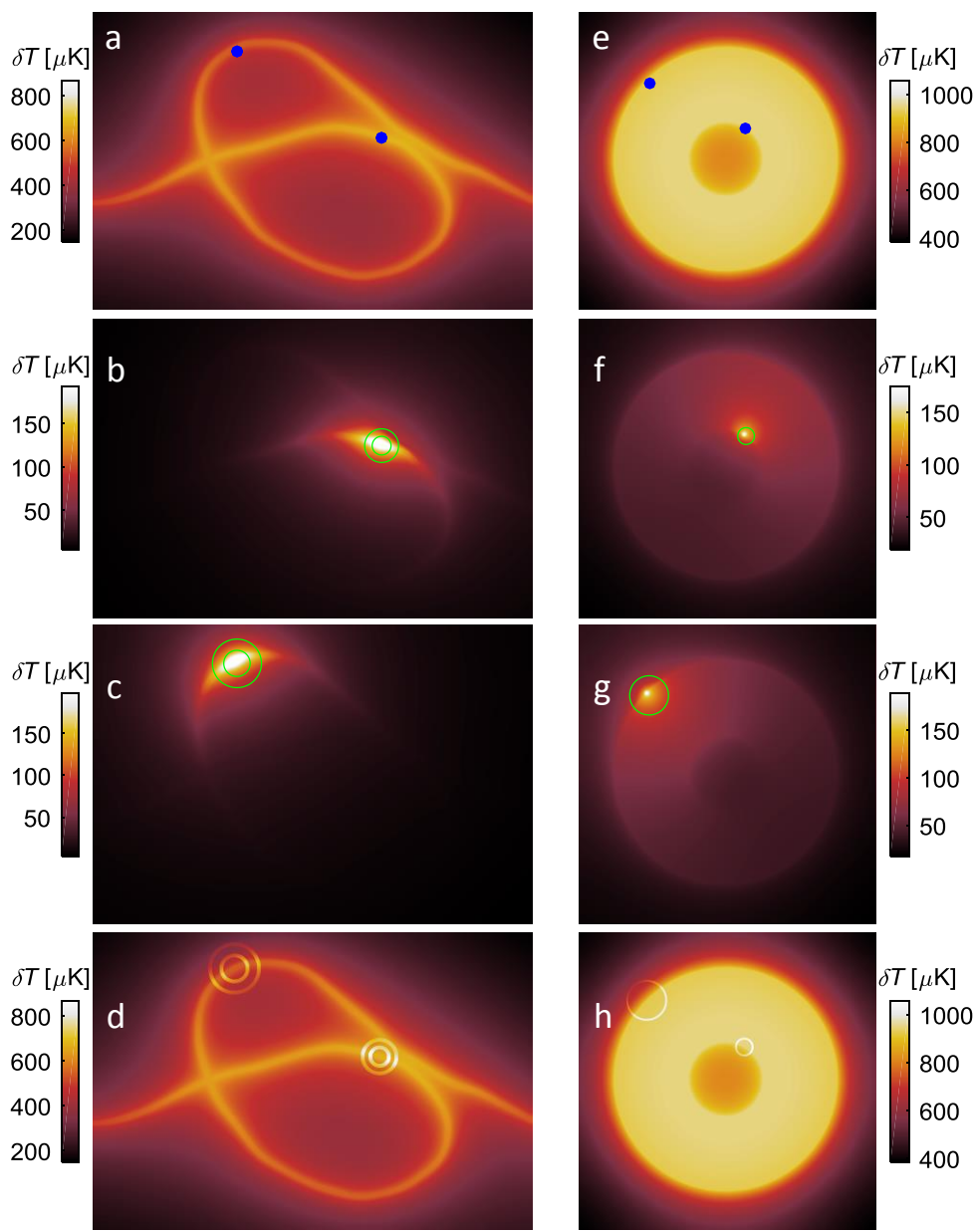

**Figure S13 | Numerical simulations of thermal ring-like structures due to resonant states in CNT and graphene. a,** Surface temperature distribution $\delta T(x,y)$ due to 20 pW/µm power dissipation in a looped CNT on a SiO$_2$ substrate. **b,** $\delta T(x,y)$ due to 20 pW dissipation at a point on the CNT marked in blue in (a). **c,** Similar $\delta T(x,y)$ due to the left point heat source marked in (a). **d,** Surface temperature distribution resulting from adding $\delta T(x,y)$ values along the green rings in (b) and (c) to $\delta T(x,y)$ of (a). **e,** Surface temperature $\delta T(x,y)$ due to 40 pW/µm$^2$ power dissipation in a washer shaped graphene structure on a SiO$_2$ substrate. **f,** $\delta T(x,y)$ due to 60 pW dissipation at a point in the graphene marked in blue in (e). **g,** Similar $\delta T(x,y)$ due to the heat source at left point marked in (e). **h,** Surface temperature attained by adding $\delta T(x,y)$ values along the green rings in (f) and (g) to $\delta T(x,y)$ of (e).

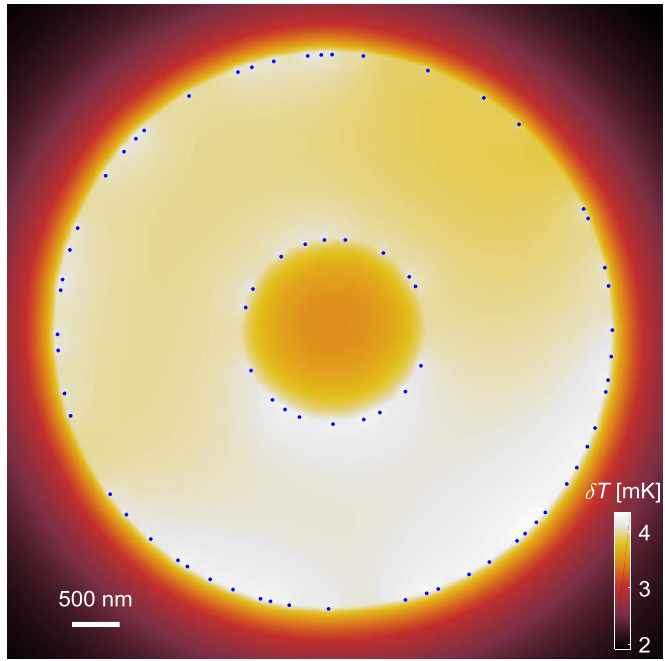

**Figure S14 | Numerical simulation of an ensemble of point heat sources in graphene.** Simulation of the surface temperature $\delta T(x,y)$ due to random distribution of point heat sources along the washer-shaped graphene edges on $SiO_2$ substrate. The locations of the sources are marked by blue dots with an average density of 3 points/μm, which is sparser compared to the experimental findings. A power of 60 pW is dissipated at each point, resulting in a smeared image in which the ability to resolve the individual heat sources is lost, in contrast to Fig. 3.

Even though the ring structures may seem like a measurement artifact, they in fact provide a powerful spectroscopic tool that allows detection of the QDs or other resonant dissipative states which would not be possible otherwise. Figure S14 shows the calculated surface temperature distribution of the graphene washer similar to Fig. S13e in which instead of a uniform bulk heating all the dissipation arises due to isolated point heat sources, like in Figs. S13f and g, randomly distributed along the edges with even sparser average density (3 sources/µm) as compared to the experimental findings of Fig. 3. The resulting $\delta T(x, y)$ shows smooth inhomogeneities due to variations in the local average density of the point sources, smeared by the long-range heat conductivity, with essentially no visible signature of the individual sources. The scanning gate ability of the tSOT, in contrast, allows controllable switching of each of the point sources thus providing a unique tool for their individual detection and characterization.

## S14. Scanning gate thermometry of QD in SWCNT

In addition to the scanning *dc* gate thermometry demonstrated in Fig. 2 in which the potential of the tSOT was kept constant while an *ac* current was applied to the CNT, we have performed scanning *ac* gate thermometry with an *ac* voltage of 80 mV at 130 Hz applied to the tSOT, while a *dc* current of 1.7 nA flowed through the CNT device 2. In this case the heat signal at the excitation frequency is generated only locally due to the small induced changes in the local conductance, reducing the background signal due to the overall *ac* heating of the CNT observed in Figs. 2a and 2b. Figures S15a-c present the $T_{ac}$ images produced by this technique at three different heights $h$ of the tSOT above a CNT segment marked by square in Fig. 2e. The height $h$ was incremented by 2 nm between the images. As expected, the diameter of the Coulomb blockade rings of the QDs grows with $h$. Clearly the ring structure is far from circular. In all cases, the short axis of the asphericity of the rings is aligned with the direction of the CNT; a feature that may be explained by partial screening of the tSOT potential by neighboring QDs along the CNT. In addition, during the scan of Fig. S15b an abrupt change occurred, causing some rings to shrink by one period while other rings remained unchanged (scans are along vertical lines from right to left). This observation reveals instability, typical for QDs, in which the charge of one QD is abruptly changed without affecting the rest of the QDs, demonstrating the ability of scanning gate thermometry to resolve changes in dissipation arising from variations in single electron occupancy of a QD.

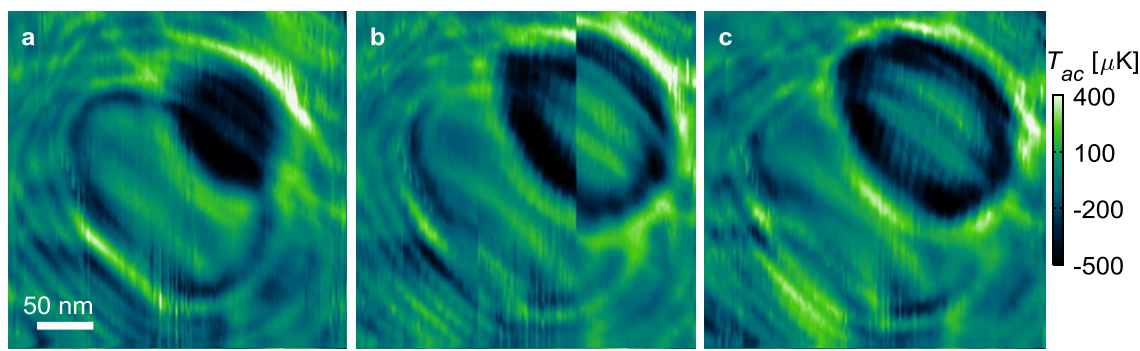

**Figure S15 | Thermal imaging by scanning *ac* gate thermometry of quantum dot dissipation in SWCNT. a-c,** $T_{ac}$ images of a segment along CNT device 2 (marked by square in Fig. 2e) arising from changes in QDs dissipation induced by an *ac* potential applied to the tSOT in presence of *dc* current of 1.7 nA in the CNT. The height of the tSOT above the CNT was increased by 2 nm between the consecutive images, starting from about 15 nm in (a), resulting in expansion of the Coulomb blockade rings. During scan (b) an abrupt jump in the charge occupancy of one of the QDs occurred without affecting the dissipation in other QDs. Pixel size 1.8 nm and scan speed 10 nm/s.

## S15. Scanning gate thermometry of hBN/graphene/hBN device

We employed three schemes for scanning gate thermometry of the graphene device, one *dc* scheme and two additional *ac* schemes introduced to enhance contrast. In the simplest scheme a *dc* current of 6 µA is driven through the structure and the corresponding $\delta T_{dc}$ image is acquired by the tSOT as shown in Fig. S16a, which presents a zoomed-in view of the inner aperture of the washer-shaped device. The applied current raises the temperature in the graphene region relative to the inner graphene-free region outlined by the red dotted circle. In addition to the smooth temperature background in graphene, a neckless of small ring-like structures of enhanced temperature with a radius of up to about $R = 150$ nm is clearly visible along the edge of the graphene. Since the temperature increase at the rings is relatively small compared to the background temperature we employ two schemes in order to enhance their visibility.

In the first *ac*–enhanced imaging mode we make use of the fact that the tSOT is mounted on a tuning fork (TF, see section S8 for more information). We excite the TF at its resonant frequency of 36.694 kHz with an rms amplitude of $|\vec{u}|$ =5 nm and measure the resulting *ac* signal using a lock-in amplifier. The *ac* displacement of the tSOT gives rise to an *ac* signal $T_{ac} = \vec{u} \cdot \nabla T_{dc}$ which reflects the gradient of $\delta T_{dc}(x, y)$ along $\hat{u}$ – the direction of the TF oscillations, that were oriented at 30 deg with respect to the $x$ axis, as indicated by the arrow in Fig. S16b. In the resulting $\hat{u} \cdot \nabla T_{dc}(x, y)$ gradient image in Fig. S16b, which was acquired simultaneously with the $\delta T_{dc}(x, y)$ image in Fig. S16a, the smoothly varying background is suppressed, while the sharp ring-like features are enhanced, thus facilitating a more detailed analysis of their behavior. The obtained image exhibits red and blue features due to positive and negative temperature gradient at the inner and outer sides of the *dc* rings.

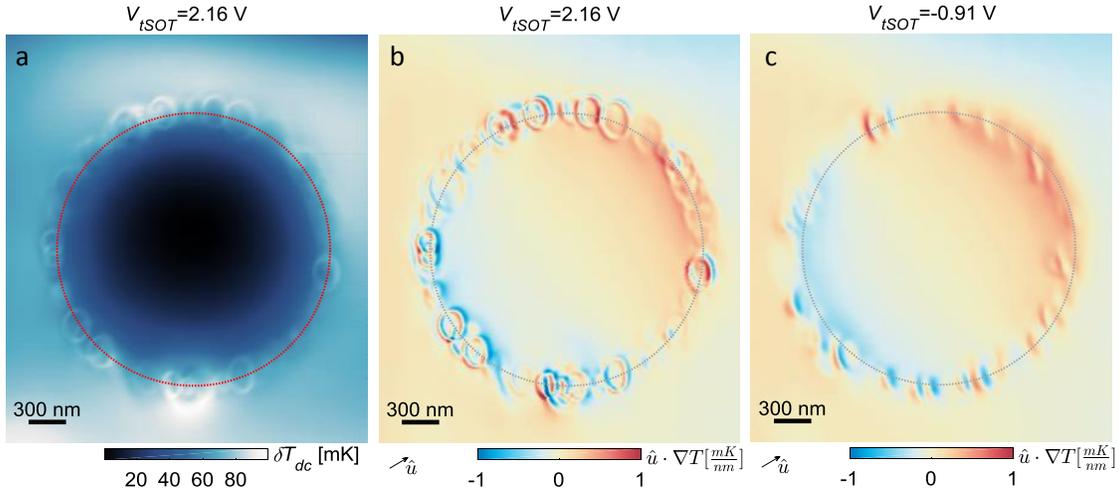

**Figure S16 | Evolution of circular scanning gate thermometry signatures of resonant dissipation in graphene.** The Back-gate of the graphene device of Fig. 3 was tuned to a fixed carrier concentration of $1.5 \cdot 10^{12}$ cm$^{-2}$. A *dc* current of 6 µA flows from top to bottom similarly to Fig. 3. The tSOT scans 20 nm above the structure to image the inner hole of the structure presented in Fig. 3. The tSOT gating voltage was swept to create a series of thermal maps. **a,** A typical thermal map for a given tSOT gating voltage. The warm bright rings on the edge of the structure are clearly visible. **b,** When locked to the tuning-fork frequency of 36.694 kHz, one simultaneously gets a spatial derivative of the thermal map along the vibration axis (indicated by a unit vector u in the images) with an improved contrast. Two representing tSOT gate voltages are presented (b-c). The structure edge is marked by a dashed line. A full video of the evolution of the patterns is provided as well (Video S1). Pixel size 18.7 nm and time constant 3 ms.

Using this *ac*-enhanced imaging mode we obtain a behavior of the rings with varying the tip-sample bias voltage $V_{tSOT}$, as illustrated in Figs. S16b,c and Video S1. The back-gate voltage was fixed at -2.5 V yielding carrier concentration of $1.5 \cdot 10^{12}$ cm$^{-2}$ and a series of images were acquired at different values of $V_{tSOT}$ using a 100 nm diameter tSOT scanning at a height of 20 nm above the hBN/graphene/hBN structure. The Video S1 shows that upon increasing $V_{tSOT}$ rings start to appear at various locations along the edge of graphene and their radii grow as the tip bias increases consistent with the numerical simulations in section S13.

As described in the main text, we attribute this behavior to the presence of localized electronic resonance states at adatoms and vacancies positioned at the exposed etched edges of the graphene. Recent studies have shown that these defects can indeed give rise to localized resonant states that are pinned in energy close to the Dirac point[32,54-56]. When a negative back gate is applied these localized states reside well above the Fermi energy and therefore cannot trap the hot carriers that have characteristic energies only slightly above $E_F$. By placing the tip directly above the defect and increasing $V_{tSOT}$ the energy level of the localized state is shifted towards $E_F$ and comes into resonance with the hot electrons at $V_{tSOT} = V_c$. At this bias value a heat spot will be formed at the defect below the tip as a result of electron-lattice cooling bottleneck relieved through the defect resonances. Under these conditions the hot carriers, trapped on the localized resonance states, release their excess energy through electron-phonon coupling enhanced by the localized states.

As $V_{tSOT}$ is further increased the energy level of the localized state is shifted to below $E_F$ and thus goes out of resonance with hot electrons. However, in this situation where $V_{tSOT} > V_c$, the resonant conditions of the defect state can be restored by reducing the tip-induced potential by moving the tip a distance $R$ away from the defect. In this case a ring of enhanced temperature of radius $R$ in the $\delta T_{dc}(x,y)$ images will be observed as in Fig. S16. Additional increase in $V_{tSOT}$ requires moving the tip further away from the defect. Hence, the radius $R$ of the rings grows with $V_{tSOT}$ as shown in Video S1. The value of $V_c$ should depend on the specific structure of the defect and its environment and is hence expected to be location dependent. As a result, the onset of the ring formation should vary from one defect to another. Thus for a given $V_{tSOT}$ a distribution of ring sizes should be observed, consistent with our experimental findings. Nevertheless, the various defects show similar behavior indicating a common nature of the localized states.

Since the ring radius $R$ grows with $V_{tSOT}$, we can use an alternative *ac*-enhanced mode to improve the ring contrast. Adding a small *ac* component to $V_{tSOT}$ will result in a small periodic expansion and shrinkage of the ring giving rise to a $T_{ac}(x,y)$ image in which the rings appear with bright and dark outlines. Figure 3b was acquired using this scheme of scanning gate thermometry with tip *ac* voltage of 200 mV at 843 Hz. The results obtained by all three scanning modes are found to be in good agreement with each other.

## S16. Supplementary Video S1: Scanning gate thermometry of hBN/graphene/hBN device

Video S1 presents a sequence of *ac* images of the inner aperture of the washer-shaped graphene device acquired by the tSOT as a function of the potential $V_{tSOT}$ between the tip and the sample. The images present the gradient of the local temperature $\hat{u} \cdot \nabla T_{dc}(x,y)$ produced by measuring the *ac* signal of the tSOT due to the *ac* vibration of the TF with an rms amplitude of 5 nm along $\hat{u}$ at 36.694 kHz as described in section S16 above. Figures S16b,c present two frames from the video. New ring-like structures are formed along the graphene edge and expand as $V_{tSOT}$ is increased as discussed in sections S13 and S16.